\documentclass[aps,pre,tightenlines,graphicx]{revtex4}

\usepackage[dvips]{graphicx}
\input{psfig.sty}

\newcommand{\la}{\left\langle}
\newcommand{\ra}{\right\rangle}
\newcommand{\EPL}{{\it Europhys.~Lett.~}}
\newcommand{\PRL}{{\it Phys.~Rev.~Lett.~}}
\newcommand{\PR}{{\it Phys.~Rev.~}}
\newcommand{\JCP}{{\it J.~Chem.~Phys.~}}
\newcommand{\JPC}{{\it J.~Phys.~Chem.~}}
\newcommand{\JSP}{{\it J.~Stat.~Phys.~}}
\newcommand{\JPCM}{{\it J.~Phys.: Condens.~Matter~}}
\newcommand{\MP}{{\it Mol.~Phys.~}}
\newcommand{\JCIS}{{\it J.~Coll.~Int.~Sci.~}}
\newcommand{\EPJ}{{\it Eur.~Phys.~J.~}}

\begin{document}

 
\title{Nonlinear Screening and Effective Electrostatic Interactions 
in Charge-Stabilized Colloidal Suspensions}

\author{A. R. Denton}
\email{alan.denton@ndsu.nodak.edu}
\affiliation{Department of Physics, North Dakota State University,
Fargo, ND, 58105-5566}

\date{\today}
\begin{abstract}
A nonlinear response theory is developed and applied to electrostatic 
interactions between spherical macroions, screened by surrounding microions, 
in charge-stabilized colloidal suspensions.  The theory describes 
leading-order nonlinear response of the microions (counterions, salt ions) 
to the electrostatic potential of the macroions and predicts microion-induced 
effective many-body interactions between macroions.  
A linear response approximation [{\it Phys.~Rev.} E {\bf 62}, 3855 (2000)] 
yields an effective pair potential of screened-Coulomb (Yukawa) form,
as well as a one-body volume energy, which contributes to the free energy.
Nonlinear response generates effective many-body interactions and essential 
corrections to both the effective pair potential and the volume energy.  
By adopting a random-phase approximation (RPA) for the response functions, 
and thus neglecting microion correlations, practical expressions are 
derived for the effective pair and triplet potentials 
and for the volume energy.
Nonlinear screening is found to weaken repulsive pair interactions, 
induce attractive triplet interactions, and modify the volume energy.
Numerical results for monovalent microions are in good agreement with 
available {\it ab initio} simulation data and demonstrate that 
nonlinear effects grow with increasing macroion charge 
and concentration and with decreasing salt concentration.  
In the dilute limit of zero macroion concentration, leading-order 
nonlinear corrections vanish.  Finally, it is shown that 
nonlinear response theory, when combined with the RPA, 
is formally equivalent to the mean-field Poisson-Boltzmann theory and that
the linear response approximation corresponds,
within integral-equation theory, to a linearized hypernetted-chain closure.
\end{abstract}

\pacs{82.70.Dd, 83.70.Hq, 05.20.Jj, 05.70.-a}

\maketitle




\section{Introduction}

Electrostatic interactions between charged macromolecules dispersed in an 
electrolyte solvent have attracted sustained cross-disciplinary interest 
because of their fundamental role in governing the physical properties of 
colloidal suspensions~\cite{Hunter,Pusey,Schmitz}, polyelectrolyte 
solutions~\cite{PE1,PE2}, and many biological systems. 
Colloids (nm-$\mu$m-sized particles) and polyelectrolytes 
(charged polymers) can acquire charge in solution through dissociation 
of counterions.  
Familiar examples of charged colloids are latex or silica microspheres, 
clay platelets, and ionic micelles suspended in water.
Common polyelectrolytes are polyacrylic acid, found in gels and 
rheology modifiers, and biopolymers ({\it e.g.}, DNA, proteins, starches) 
in aqueous solution.
In all of these systems, bare Coulomb interactions between charged 
macromolecules (macroions) are screened by counterions and salt ions 
(microions).
This paper formulates a general response theory of microion screening and 
applies the theory to microion-induced effective pair and many-body 
interactions between colloidal macroions in suspension.

In recent years, experimental reports of apparent attractions between 
like-charged macroions have focused attention on electrostatic 
interactions in strongly charged, deionized suspensions.  
Observations of anomalous thermodynamic behavior, 
such as bulk phase separation~\cite{Tata,Ise,Matsuoka,Groehn00} and
metastable crystallites~\cite{Grier2}, and direct measurements of 
attractive interactions between confined macroions~\cite{Fraden,Grier1}
have motivated a variety of theories and computer simulation methods, which 
have been recently reviewed~\cite{Levin02,Likos01,Belloni00,Hansen-Lowen}.

A variety of simulation methods have been applied to explore effective 
interparticle interactions and phase behavior in charged colloids.  
Standard molecular dynamics~\cite{Robbins88} and 
Monte Carlo~\cite{Meijer91,Auer02,Hynninen03} algorithms have been used to 
investigate crystallization in effective one-component pairwise-interacting 
systems, while powerful {\it ab initio} 
(classical Car-Parrinello)~\cite{Lowen,Tehver} and multi-component 
Monte Carlo~\cite{Stevens96,Linse-Lobaskin,Holm,Damico} techniques have 
modeled effective interactions and, to a lesser degree, phase behavior.

Theoretical approaches can be broadly distinguished by the extent to which 
they include correlations between microions. 
Many approaches are founded on the Poisson-Boltzmann (PB) equation
for the electrostatic potential, which is derived from mean-field 
approximations that neglect microion correlations.
The classic theory of Derjaguin and Landau~\cite{DL} and Verwey and 
Overbeek~\cite{VO} (DLVO), based on a linearization of the PB equation, 
predicts that widely separated macroions interact via a purely repulsive 
effective electrostatic pair potential of screened-Coulomb (Yukawa) form.  
Similar effective interactions have been derived within the 
frameworks of density-functional (DF) theory~\cite{Lowen,Graf,vRH,vRDH},
response theory~\cite{Silbert,Denton1,Denton2}, and extended Debye-H\"uckel 
(DH) theories~\cite{Warren,Chan01}.  These more recent approaches 
also clarify the importance of a one-body volume 
energy~\cite{Graf,vRH,vRDH,Silbert,Denton1,Denton2,Warren,Chan01},
which contributes a state-dependent term to the free energy and thus 
can influence thermodynamic behavior.

Microion correlations, while often weak for monovalent microions, generally 
cannot be ignored in the case of multivalent microions, as emphasized 
in several recent studies of charged colloids and 
polyelectrolytes~\cite{Holm,Rouzina96,Liu97,Stevens99,Shklovskii99,Nguyen00}.
Microion correlations can induce short-range attractions, which have 
been linked to condensation of DNA and other 
polyelectrolytes~\cite{Rouzina96,Liu97,Stevens99,Gelbart01}.
Another wide class of theories that include some microion correlations
is the class of integral-equation theories~\cite{Patey80,Belloni86,
Khan87,Outhwaite02,Carbajal-Tinoco02, Petris02,Anta}, which predict 
multi-component correlation functions from the Ornstein-Zernike relation 
combined with various closures.

Many theoretical approaches rely, in practice, on some manner of linear 
approximation.  DLVO theory and linearized PB 
cell models~\cite{vonGrunberg01,Deserno02,Tamashiro03} are based on 
the linearized PB equation.  The DF~\cite{Graf,vRH,vRDH} and response 
theory~\cite{Silbert,Denton1,Denton2} approaches involve truncating 
expansions of free energy functionals or of microion density profiles.
While linear approximations can be justified under a wide range of 
conditions, their validity may be questioned for concentrated suspensions 
of highly charged macroions at low salt concentrations (ionic strengths) 
-- precisely those conditions under which anomalous phase behavior has 
been reported.  On the other hand, many nonlinear theories, such as 
the full PB theory~\cite{Bowen-Sharif98,Gray99,Dobnikar03} and
integral-equation theories, present severe computational challenges.
In fact, the nonlinear PB equation usually yields to numerical solution 
only within cell models with simplified boundary conditions.

The main purpose of the present paper is to extend response theory to include 
leading-order nonlinear microion screening and to apply the extended theory 
to systematically test the linear-screening approximation.  This extension
necessarily entails three-body effective interactions between macroions and 
corrections at the pair and one-body levels, for which computationally 
practical expressions are derived.  The predicted effective interactions 
could, in future studies, be input directly into statistical mechanical 
theories or simulations to study influences of nonlinear screening 
on phase equilibria and other phenomena.  

The key qualitative conclusion of the paper is 
that nonlinear effects can significantly modify effective interactions,
becoming increasingly important with increasing macroion charge and 
concentration and with decreasing salt concentration.  
Numerical calculations for bulk suspensions are performed to quantify 
parameter ranges wherein linearization is justified.  Comparison is made
with a similar extension of the DF approach, recently applied to 
wall-induced effective pair interactions~\cite{Goulding99,HGvR00} 
and to effective triplet interactions~\cite{LA}.

Outlining the remainder of the paper, Sec.~\ref{Model} defines the model 
colloidal suspension; Sec.~\ref{Theory} develops a general response theory 
for the system; Sec.~\ref{Results} presents analytical results for 
leading-order nonlinear corrections to the effective microion-induced 
interactions; Sec.~\ref{Numerical} presents numerical results, for selected 
parameters, and comparisons with predictions of linear response theory; 
Sec.~\ref{Conclusions} summarizes the paper; and finally the Appendix 
compares response theory with two related approaches, namely PB theory 
and integral-equation theory.

\section{Model}\label{Model}
The system of interest comprises colloidal macroions, counterions, and 
salt ions dispersed in a solvent (Fig.~\ref{fig-model}).  
This multi-component mixture is modeled here as a collection of $N_m$ 
charged hard-sphere macroions, of valence $-Z$ (surface charge $-Ze$) 
and radius $a$ (diameter $\sigma=2a$), and 
$N_c$ point counterions of valence $z$ in an electrolyte solvent of
volume $V$ at temperature $T$.  Global charge neutrality constrains 
macroion and counterion numbers via the relation $ZN_m=zN_c$.
For simplicity, we assume a symmetric electrolyte 
consisting of $N_s$ point salt ions of valence $z$ and $N_s$ of valence $-z$ 
(same valence as counterions) in a uniform solvent.  The microions thus 
number $N_+=N_c+N_s$ positive and $N_-=N_s$ negative, for a total of 
$N_{\mu}=N_c+2N_s$.  
The solvent is approximated, within the primitive model, as a dielectric 
continuum, characterized entirely by a dielectric constant $\epsilon$.  

The macroion charge, which may be physically interpreted as an effective
(renormalized) charge, is assumed to be fixed and distributed smoothly 
over the particle surface.  Charge discreteness can be reasonably neglected
if the distance between neighboring macroion surfaces much exceeds
the typical distance between charge groups on a macroion surface,
roughly $\sigma/\sqrt{Z}$.  
The assumption of point microions limits the model to systems with large 
size asymmetries.  Furthermore, we neglect polarization effects, {\it e.g.},
charge-induced dipole interactions~\cite{Fisher94,Phillies95,Gonzalez01},
which are shorter-ranged than charge-charge interactions, and which vanish if 
solvent and macroions have the same dielectric constant 
({\it i.e.}, are index matched).

\section{Theory}\label{Theory}
\subsection{Reduction to One Component}
The response theory of effective interactions is fundamentally based 
on a reduction of the multi-component mixture to 
an equivalent one-component system by integrating out the degrees of freedom
of the microions~\cite{Rowlinson84}.  In this reduction, the macroions are 
regarded as applying an ``external" potential that perturbs the (otherwise 
uniform) microion distribution.  For a sufficiently weak perturbation 
(dilute or weakly charged macroions), the microions respond linearly.  
The linear response approximation has been discussed in 
refs.~\cite{Silbert,Denton1,Denton2}.
Upon increasing the macroion charge or concentration, however,
nonlinear microion response becomes increasingly important.
This motivates the current extension of response theory from linear to 
nonlinear response.  

To simplify the derivation, we first consider salt-free suspensions and 
introduce salt ions only at the end.  The model system is then 
described by a Hamiltonian $H$ that decomposes naturally into three terms:
\begin{equation}
H=H_m(\{{\bf R}\})+H_c(\{{\bf r}\})+H_{mc}(\{{\bf R}\},\{{\bf r}\}), 
\label{H}
\end{equation}
where $\{{\bf R}\}$ and $\{{\bf r}\}$ denote the coordinates of
macroions and microions, respectively. The first term on the right
side of Eq.~(\ref{H}) is the bare macroion Hamiltonian, given by
\begin{equation}
H_m=H_{\rm HS}(\{{\bf R}\})+
\frac{1}{2}\sum_{{i\neq j=1}}^{N_m} v_{mm}(|{\bf R}_i-{\bf R}_j|),
\label{Hm1}
\end{equation}
where $H_{\rm HS}$ is the Hamiltonian for neutral hard spheres 
(the macroion hard cores) and $v_{mm}(r)=Z^2e^2/\epsilon r$, $r>\sigma$, is 
the bare Coulomb pair interaction between macroions.  In the primitive model, 
the solvent acts only to reduce the strength of Coulomb interactions 
by a factor $1/\epsilon$.  The second term of the Hamiltonian,
\begin{equation}
H_c=K_c+\frac{1}{2}\sum_{{i\neq j=1}}^{N_c} v_{cc}(|{\bf r}_i-{\bf r}_j|), 
\label{Hc1}
\end{equation}
describes the counterions alone, having kinetic energy $K_c$ and interacting 
via a Coulomb pair potential $v_{cc}(r)=z^2e^2/\epsilon r$.
The third term is the macroion-counterion interaction energy:
\begin{equation}
H_{mc}=\sum_{i=1}^{N_m}\sum_{j=1}^{N_c} v_{mc}(|{\bf R}_i-{\bf r}_j|), 
\label{Hmc1}
\end{equation}
where $v_{mc}(r)$ is the macroion-counterion electrostatic pair interaction:
$v_{mc}(r)=Zze^2/\epsilon r$, $r>a$.  For impenetrable hard-core 
macroions, the form of $v_{mc}(r)$ inside the core is arbitrary and can
be specified so as to minimize counterion penetration inside the cores
(see Sec.~\ref{Counterion Density}).
For later reference, we note that the macroion and counterion Hamiltonians
[Eqs.~(\ref{Hm1}) and (\ref{Hc1})] may be expressed in terms of Fourier
components using the identity
\begin{equation}
\sum_{{i\neq j=1}}^N v(|{\bf r}_i-{\bf r}_j|)
=\frac{1}{V}\sum_{\bf k}\hat v(k) \left[\hat\rho({\bf k})
\hat\rho(-{\bf k})-N\right], \label{Hidentity}
\end{equation}
where $\hat v(k)$ is the Fourier transform of a pair potential $v(r)$, 
$\hat\rho({\bf k})$ is the Fourier transform of the appropriate 
(macroion or counterion) number density operator
$\rho({\bf r})=\sum_{i=1}^N\delta({\bf r}-{\bf r}_i)$, and the Fourier 
transform convention is
\begin{equation}
\hat\rho({\bf k})=\int{\rm d}{\bf r}\,\rho({\bf r})
e^{-i{\bf k}\cdot{\bf r}}, \label{FTa}
\end{equation}
\begin{equation}
\rho({\bf r})=\frac{1}{V}\sum_{\bf k}\hat\rho({\bf k})
e^{i{\bf k}\cdot{\bf r}}. \label{FTb}
\end{equation}
The inverse transform is expressed as a summation, rather than an integral,
in anticipation that charge neutrality will necessitate singling out the 
$k=0$ component for special treatment.

At constant temperature and volume, the thermodynamic behavior of the system 
is governed by the canonical partition function,
\begin{equation}
{\cal Z}=\la\la\exp(-\beta H)\ra_c\ra_m, 
\label{part1}
\end{equation}
where $\beta=1/k_BT$ and $\la~\ra_c$ and $\la~\ra_m$ denote classical 
traces over counterion and macroion coordinates, respectively. 
The two-component mixture of macroions and counterions can be
formally mapped onto an equivalent one-component system of
``pseudo-macroions" by performing a restricted trace over
counterion coordinates, keeping the macroions fixed.  Thus, without
approximation,
\begin{equation}
{\cal Z}=\la\exp(-\beta H_{\rm eff})\ra_m, 
\label{part2}
\end{equation}
where $H_{\rm eff}=H_m+F_c$ is the effective Hamiltonian of the 
equivalent one-component system and 
\begin{equation}
F_c=-k_BT\ln\la\exp\left[-\beta(H_c+H_{mc})\right]\ra_c
\label{Fc1}
\end{equation}
can be interpreted as the free energy of a nonuniform gas of
counterions in the presence of the fixed macroions. 
To simplify notation, we henceforth omit the subscript $c$ from the trace
over counterion coordinates: $\la~\ra_c\equiv \la~\ra$.

\subsection{Response Theory}\label{Response Theory}
Although the one-component mapping is exact, the challenge now shifts 
to approximating the counterion free energy $F_c$. 
Progress can be made by regarding the macroions as an ``external" potential 
for the counterions and invoking perturbation theory~\cite{Silbert,HM}:
\begin{equation}
F_c=F_0+\int_0^1{\rm d}\lambda\,\la H_{mc}\ra_{\lambda},
\label{Fc2}
\end{equation}
where $F_0=-k_BT\ln\la\exp(-\beta H_c)\ra$ is the reference free energy 
of the counterions in the presence of neutral (hard-core) macroions 
(the counterions then being unperturbed, except for exclusion from
the macroion cores), $\la~\ra_{\lambda}$ denotes a trace over
coordinates of the counterions in the presence of macroions charged
to a fraction $\lambda$ of their full charge, and the $\lambda$-integral 
adiabatically charges the macroions from neutral to fully charged.  
Although each term on the right side of Eq.~(\ref{Fc2}) is infinite,
the infinities cancel to yield a finite counterion free energy.
When the macroions are uncharged, the surrounding ``sea" of counterions 
has uniform density, neglecting any confinement-induced
structure, which is reasonable for typical counterion concentrations in 
colloidal suspensions (see below).  As the macroion charge is ``turned on," 
the counterions respond, redistributing themselves to form a double layer 
(surface charge plus neighboring counterions) surrounding each macroion. 

In practice, it proves convenient to convert $F_0$ to the free energy
of a classical one-component plasma (OCP) by adding and subtracting 
the energy of a uniform compensating negative background.
The background energy can be expressed as 
$E_b=-\frac{1}{2}N_cn_c\hat v_{cc}(0)$, where $n_c$ is the average density
of counterions in the volume unoccupied from the macroion cores.  
Note that the infinite background energy formally cancels the infinities 
on the right side of Eq.~(\ref{Fc2}).
Because the counterions are excluded (with the background) from the hard
macroion cores, the OCP has average density $n_c=N_c/V'$, where 
$\eta=\frac{\pi}{6}(N_m/V)\sigma^3$ is the macroion volume fraction and
$V'=V(1-\eta)$ is the free volume.  Thus,
\begin{equation}
F_c=F_{\rm OCP}+\int_0^1{\rm d}\lambda\,\la H_{mc}\ra_{\lambda}-E_b,
\label{Fc3}
\end{equation}
where $F_{\rm OCP}=F_0+E_b$ is the free energy of the OCP in the presence 
of neutral, but volume-excluding, hard spheres -- what might be loosely
called a ``Swiss cheese" OCP.  

In terms of Fourier components, the macroion-counterion interaction 
can be expressed as
\begin{equation}
\la H_{mc}\ra_{\lambda}=\frac{1}{V'}\sum_{\bf k} \hat v_{mc}(k) 
\hat\rho_m({\bf k}) \la\hat\rho_c(-{\bf k})\ra_{\lambda}.
\label{Hmc2}
\end{equation}
Evidently, $\la H_{mc}\ra_{\lambda}$ depends through 
$\la\hat\rho_c ({\bf k})\ra_{\lambda}$
upon the response of the counterions to the macroion charge density. 
Note, however, that there is no response for $k=0$, since
$\hat\rho_c(0)=\int{\rm d}{\bf r}\,\rho_c({\bf r})=N_c$, which is fixed 
by charge neutrality for a given macroion concentration.  Taking this 
$k\to 0$ limit into account, and subtracting the background energy,
Eq.~(\ref{Hmc2}) becomes
\begin{equation}
\la H_{mc}\ra_{\lambda}-E_b=\frac{1}{V'}\sum_{\bf k\neq 0}\hat
v_{mc}(k) \hat\rho_m({\bf k})
\la\hat\rho_c(-{\bf k})\ra_{\lambda}
+n_c\lim_{k\to 0}\left[N_m\hat v_{mc}(k)+\frac{N_c}{2}
\hat v_{cc}(k)\right]. \label{Hmc3}
\end{equation}

To proceed further, we apply a perturbative approximation for the 
macroion-induced counterion density, adapting a standard approach 
from the theory of metals~\cite{HM,AS,Hafner87,Ashcroft66}.
Defining the macroion external potential $\phi_{\rm ext}({\bf r})$ by
\begin{equation}
ze\phi_{\rm ext}({\bf r})=\int{\rm d}{\bf r}'\, v_{mc}(|{\bf r}-{\bf r}'|) 
\rho_m({\bf r}'),
\label{phiext}
\end{equation}
the ensemble-averaged induced counterion density may be expanded in a 
functional Taylor series around $\phi_{\rm ext}({\bf r})=0$~\cite{note1}
in powers of the dimensionless potential 
$u({\bf r})=-\beta ze\phi_{\rm ext}({\bf r})$:
\begin{equation}
\la\rho_c({\bf r})\ra=\rho_0+
\sum_{n=1}^{\infty}\frac{1}{n!}\int{\rm d}{\bf r}_1\,
\cdots\int{\rm d}{\bf r}_n\, 
G^{(n+1)}({\bf r}-{\bf r}_1,\ldots,{\bf r}-{\bf r}_n)
u({\bf r}_1)\cdots u({\bf r}_n).
\label{delta-rho}
\end{equation}
Here $\rho_0$ is a constant, chosen below to ensure charge neutrality,
and the coefficients
\begin{equation}
G^{(n+1)}({\bf r}-{\bf r}_1,\ldots,{\bf r}-{\bf r}_n)=\lim_{u\to 0}
\left(\frac{\delta^n\la\rho_c({\bf r})\ra}
{\delta u({\bf r}_1)\cdots\delta u({\bf r}_n)}\right)
\label{G}
\end{equation}
are the $(n+1)$-particle density correlation functions~\cite{HM} of the 
unperturbed (uniform) OCP.
The correlation functions are, in turn, proportional to response functions
(see below).  To give a physical interpretation to Eq.~(\ref{delta-rho}),
the counterion density induced at any point ${\bf r}$ is the net response
to macroion-generated external potentials, applied at sets of points 
$\{{\bf r}_1,\ldots,{\bf r}_n\}$, and propagated through the OCP via 
multi-particle density correlations.
Fourier transforming Eq.~(\ref{delta-rho}), we obtain (for $k\neq 0$)
\begin{eqnarray}
\la\hat\rho_c({\bf k})\ra~&=&~\hat G^{(2)}(k)\hat u({\bf k})
+\frac{1}{2V'}\sum_{{\bf k}'}\hat G^{(3)}({\bf k}',{\bf k}
-{\bf k}')\hat u({\bf k}')\hat u({\bf k}-{\bf k}') \nonumber
\\ ~&+&~ \frac{1}{3!V'^2}\sum_{{\bf k}',{\bf k}''}\hat G^{(4)}
({\bf k}',{\bf k}'',{\bf k}-{\bf k}'-{\bf k}'')\hat u({\bf k}')
\hat u({\bf k}'')\hat u({\bf k}-{\bf k}'-{\bf k}'')+\cdots. \label{rhock1}
\end{eqnarray}
The coefficients $\hat G^{(n)}$, which are Fourier transforms of $G^{(n)}$, 
are related to the $n$-particle static structure factors of the 
uniform OCP via $\hat G^{(n)}=n_cS^{(n)}$, where
the static structure factors are explicitly defined by~\cite{HM}
\begin{equation}
S^{(2)}(k)\equiv S(k)
=\frac{1}{N_c}\la\hat\rho_c({\bf k})\hat\rho_c(-{\bf k})\ra
\label{S2}
\end{equation}
and
\begin{equation}
S^{(n)}({\bf k}_1,\cdots,{\bf k}_{n-1})=\frac{1}{N_c}\la\hat\rho_c({\bf k}_1)
\cdots\hat\rho_c({\bf k}_{n-1})\hat\rho_c(-{\bf k}_1-\dots-{\bf k}_{n-1})\ra,
\quad n\geq 3.
\label{Sn}
\end{equation}
Substituting $\hat u({\bf k})=-\beta\hat v_{mc}(k)\hat\rho_m({\bf k})$
[from Eq.~(\ref{phiext})] into Eq.~(\ref{rhock1}), 
the induced counterion density can be expressed in the equivalent form
\begin{eqnarray}
\la\hat\rho_c({\bf k})\ra~&=&~\chi(k)
\hat v_{mc}(k)\hat\rho_m({\bf k})+\frac{1}{V'}\sum_{{\bf k}'}
\chi'({\bf k}',{\bf k}-{\bf k}')
\hat v_{mc}(k') \hat v_{mc}(|{\bf k}-{\bf k}'|) \nonumber \\
~&\times&~\hat\rho_m({\bf k}') \hat\rho_m({\bf k}-{\bf k}')
+\cdots, \qquad k\neq 0, \label{rhock2}
\end{eqnarray}
where 
\begin{equation}
\chi(k)=-\beta n_cS(k)
\label{chi1k}
\end{equation}
and
\begin{equation}
\chi'({\bf k}',{\bf k}-{\bf k}')=
(\beta^2n_c/2)S^{(3)}({\bf k}',{\bf k}-{\bf k}') 
\label{chi2k}
\end{equation}
are, respectively, the linear and the first nonlinear response function 
of the uniform OCP.
The first term on the right side of Eq.~(\ref{rhock2}) represents the 
linear response approximation -- linear in $\hat\rho_m({\bf k})$ -- while 
the higher-order terms generate, as shown below, nonlinear corrections 
to both the counterion density and the effective interactions.
Finally, since the amplitude of $\hat v_{mc}(k)$ is proportional to the 
macroion charge, then
\begin{eqnarray}
\la\hat\rho_c({\bf k})\ra_{\lambda}~&=&~\lambda\chi(k)
\hat v_{mc}(k)\hat\rho_m({\bf k})+\frac{\lambda^2}{V'}\sum_{{\bf k}'}
\chi'({\bf k}',{\bf k}-{\bf k}')
\hat v_{mc}(k') \hat v_{mc}(|{\bf k}-{\bf k}'|) \nonumber \\
~&\times&~\hat\rho_m({\bf k}') \hat\rho_m({\bf k}-{\bf k}')
+\cdots, \qquad k\neq 0. \label{rhock2-lambda}
\end{eqnarray}
Note that the response functions describe the response of the 
fully-charged OCP, and so do not depend on the coupling constant $\lambda$.

\subsection{Effective Interactions}
We are now positioned to derive formal expressions for the effective 
interactions.
Substituting Eq.~(\ref{rhock2-lambda}) into Eq.~(\ref{Hmc3}), the latter into 
Eq.~(\ref{Fc3}), and integrating over $\lambda$, we obtain the counterion
free energy to {\it third} order in the macroion density:
\begin{eqnarray}
& &F_c=F_{\rm OCP} + n_c\lim_{k\to 0}
\left[N_m\hat v_{mc}(k) + \frac{N_c}{2}\hat v_{cc}(k)\right]
+ \frac{1}{2V'}\sum_{{\bf k}\neq 0}\chi(k)\left[\hat v_{mc}(k)\right]^2
\hat\rho_m({\bf k})\hat\rho_m(-{\bf k}) \nonumber \\
&+&\frac{1}{3V'^2}\sum_{{\bf k}\neq 0}\sum_{{\bf k}'}\chi'({\bf k}',
-{\bf k}-{\bf k}')\hat v_{mc}(k)\hat v_{mc}(k')\hat v_{mc}(|{\bf k}+{\bf k}'|)
\hat\rho_m({\bf k})\hat\rho_m({\bf k}')\hat\rho_m(-{\bf k}-{\bf k}').
\label{Fc4}
\end{eqnarray}
Evidently, the linear and first nonlinear response terms in the expansion
of $\la\hat\rho_c({\bf k})\ra$ generate terms in $F_c$ that are, respectively, 
quadratic and cubic in $\hat\rho_m({\bf k})$.  These terms can be related
to effective pair and triplet interactions between macroions.
To this end, we first identify
\begin{equation}
\hat v^{(2)}_{\rm ind}(k)=\chi(k)[\hat v_{mc}(k)]^2
\label{v2indk}
\end{equation}
as the counterion-induced macroion-macroion pair interaction in the 
linear response approximation~\cite{Silbert,Denton1,Denton2}.
In passing, we note that Eq.~(\ref{v2indk}) is similar in structure
and physical interpretation to induced interactions recently derived 
from a coarse-grained HNC theory~\cite{Anta} and from a cumulant expansion 
of the counterion partition function~\cite{Trigger}.
Combining the induced interaction with the bare Coulomb interaction yields 
the linear-response prediction for the total effective pair interaction:
\begin{equation}
\hat v^{(2)}_0(k)=\hat v_{mm}(k)+\hat v^{(2)}_{\rm ind}(k).
\label{v20k}
\end{equation}

Now the term on the right side of Eq.~(\ref{Fc4}) that is second-order 
in $\hat\rho_m({\bf k})$ may be manipulated using the identity 
[from Eq.~(\ref{Hidentity})]
\begin{equation}
\sum_{{i\neq j=1}}^{N_m} v^{(2)}_{\rm ind}(|{\bf R}_i-{\bf R}_j|)
=\frac{1}{V'}\sum_{{\bf k}\neq 0} \hat v^{(2)}_{\rm ind}(k)
\hat\rho_m({\bf k})\hat\rho_m(-{\bf k})
+\frac{N_m^2}{V'}\lim_{k\to 0}\hat v^{(2)}_{\rm ind}(k)
-N_m v^{(2)}_{\rm ind}(0).
\label{Fc4-3}
\end{equation}
Similarly, identifying
\begin{equation}
\hat v^{(3)}_{\rm eff}({\bf k},{\bf k}')=2\chi'({\bf
k}',-{\bf k}-{\bf k}') \hat v_{mc}(k)\hat v_{mc}(k')\hat
v_{mc}(|{\bf k}+{\bf k}'|) 
\label{v3effk}
\end{equation}
as an effective three-body interaction, arising from nonlinear counterion 
response, and using the identity
\begin{eqnarray}
\sum_{{i\neq j\neq k=1}}^{N_m} v^{(3)}_{\rm eff}
({\bf R}_i-{\bf R}_j,{\bf R}_i-{\bf R}_k)
~&=&~\frac{1}{V'^2}
\sum_{\bf k}\sum_{{\bf k}'}\hat v^{(3)}_{\rm eff}({\bf k},{\bf k}')
[\hat\rho_m({\bf k})\hat\rho_m({\bf k}')\hat\rho_m(-{\bf k}-{\bf k}')
\nonumber \\
~&-&~3\hat\rho_m({\bf k})\hat\rho_m(-{\bf k})+2N_m],
\label{Fc4-4}
\end{eqnarray}
the final (third-order) term in Eq.~(\ref{Fc4}) may be rewritten as
\begin{eqnarray}
&&\frac{1}{3!V'^2}\sum_{\bf k}\sum_{{\bf k}'}\hat v^{(3)}_{\rm eff}
({\bf k},{\bf k}')
\hat\rho_m({\bf k})\hat\rho_m({\bf k}')\hat\rho_m(-{\bf k}-{\bf k}')
-\frac{N_m}{3!V'^2}\sum_{\bf k}\hat v^{(3)}_{\rm eff}({\bf k},0)
\hat\rho_m({\bf k})\hat\rho_m(-{\bf k}) \nonumber \\
&=&\frac{1}{3!}\sum_{i\neq j\neq k}^{N_m}v^{(3)}_{\rm eff}({\bf R}_i-{\bf R}_j,
{\bf R}_i-{\bf R}_k)+\frac{1}{2V'^2}\sum_{\bf k}\sum_{{\bf k}'}
\hat v^{(3)}_{\rm eff}({\bf k},{\bf k}')\hat\rho_m({\bf k})\hat\rho_m(-{\bf k})
\nonumber \\
&-&\frac{N_m}{3V'^2}\sum_{\bf k}\sum_{{\bf k}'}
\hat v^{(3)}_{\rm eff}({\bf k},{\bf k}')
-\frac{N_m}{3!V'^2}\sum_{\bf k}\hat v^{(3)}_{\rm eff}({\bf k},0)
\hat\rho_m({\bf k})\hat\rho_m(-{\bf k}).
\label{third-order}
\end{eqnarray}
Combining Eqs.~(\ref{Fc4}) and (\ref{third-order}), and again invoking 
the identity in Eq.~(\ref{Hidentity}), the effective Hamiltonian acquires 
the following physically intuitive structure:
\begin{equation}
H_{\rm eff}=H_{\rm HS}+\frac{1}{2}\sum_{i\neq j=1} ^{N_m}
v^{(2)}_{\rm eff}(|{\bf R}_i-{\bf R}_j|)+\frac{1}{3!}\sum_{i\neq
j\neq k=1}^{N_m} v^{(3)}_{\rm eff} ({\bf R}_i-{\bf R}_j,{\bf R}_i-{\bf
R}_k)+E, \label{Heff2}
\end{equation}
where $v^{(2)}_{\rm eff}(r)$ and $v^{(3)}_{\rm eff}({\bf r},{\bf r}')$ 
are, respectively, the counterion-induced effective pair and triplet 
interactions in real space and $E$ is a one-body volume energy. 
In Eq.~(\ref{Heff2}), the effective triplet interaction is the 
Fourier transform of Eq.~(\ref{v3effk}), while the effective pair 
interaction is the transform of
\begin{equation}
\hat v^{(2)}_{\rm eff}(k)=\hat v^{(2)}_0(k)+\Delta\hat v^{(2)}_{\rm eff}(k), 
\label{v2effk}
\end{equation}
where
\begin{equation}
\Delta\hat v^{(2)}_{\rm eff}(k)=\frac{1}{V'}\sum_{{\bf k}'}\hat
v^{(3)}_{\rm eff}({\bf k},{\bf k}')-\frac{N_m}{3V'}\hat
v^{(3)}_{\rm eff}({\bf k},0) \label{Delta-v2effk}
\end{equation}
is the first nonlinear correction to the linear response approximation.
Note that the second term on the right side of Eq.~(\ref{Delta-v2effk})
can be traced back to the requirement of charge neutrality, which
necessitated special treatment of the $k=0$ term in Eq.~(\ref{Hmc2}).

The volume energy $E$ -- a natural by-product of reduction to an 
equivalent one-component system -- has no explicit dependence on 
macroion coordinates. 
Collecting terms that are independent of macroion coordinates, 
the volume energy takes the form
\begin{equation}
E=E_0+\Delta E, \label{E1}
\end{equation}
where
\begin{equation}
E_0=F_{\rm OCP}+\frac{N_m}{2}v^{(2)}_{\rm ind}(0) 
+N_mn_c\lim_{k\to 0}\left[\hat v_{mc}(k)
-\frac{z}{2Z}\hat v^{(2)}_{\rm ind}(k)+\frac{Z}{2z}\hat v_{cc}(k)\right]
\label{E01}
\end{equation}
is the linear response approximation~\cite{Denton1,Denton2} and
\begin{equation}
\Delta E=\frac{N_m}{6V'^2}\left[\sum_{{\bf k},{\bf k}'} \hat
v^{(3)}_{\rm eff}({\bf k},{\bf k}')-N_m\sum_{{\bf k}} \hat
v^{(3)}_{\rm eff}({\bf k},0)\right] \label{Delta-E1}
\end{equation}
is the first nonlinear correction.  
On the right side of Eq.~(\ref{E01}), the second term represents the
interaction of a macroion with its own counterions.  The terms 
in square brackets on the right side of Eq.~(\ref{E01})
and the second term on the right side of Eq.~(\ref{Delta-E1}) 
originate again from the requirement of charge neutrality.
We emphasize that nonlinear counterion response generates not only 
effective many-body interactions, but also corrections to both the 
effective pair interaction and the volume energy.  
In fact, as is clear from Eqs.~(\ref{Delta-v2effk}) and (\ref{Delta-E1}), 
the nonlinear corrections to $v^{(2)}_{\rm eff}(r)$ and $E$ are
intimately related to many-body interactions.  
Note that the volume energy depends nontrivially on the mean 
macroion density, and thus can contribute significantly to the 
total free energy of the system.

\subsection{Physical Interpretation}\label{interpretation}

While the mathematical manipulations of response theory are simpler
in Fourier space, the physical interpretation of the theory is perhaps 
more transparent in real space.  In terms of real-space functions,
the induced pair interaction, in the linear response approximation, 
can be expressed [from Eq.~(\ref{v2indk})] as
\begin{equation}
v^{(2)}_{\rm ind}(r)=\int{\rm d}{\bf r}_1\,\int{\rm d}{\bf r}_2\,
\chi(|{\bf r}_1-{\bf r}_2|)v_{mc}(r_1)v_{mc}(|{\bf r_2}-{\bf r}|),
\label{v2indr}
\end{equation}
where $\chi(|{\bf r}_1-{\bf r}_2|)$ is the real-space linear response
function, which describes the change in counterion density induced at point 
${\bf r}_2$ in response to an external potential applied at ${\bf r}_1$.
Referring to Fig.~\ref{fig-diagram}, Eq.~(\ref{v2indr}) can be interpreted
as follows.  The external potential due to one macroion 
(centered at the origin in Fig.~\ref{fig-diagram})
induces at point ${\bf r}_2$ a change in counterion density 
$\int{\rm d}{\bf r}_1\, \chi(|{\bf r}_1-{\bf r}_2|) v_{mc}(r_1)$.  
This induced density, which depends (via $\chi$) on the pair density
correlation function of the intervening medium (OCP), then interacts 
with a second macroion, at displacement ${\bf r}$ from the first, giving rise 
to a counterion-induced pair interaction energy.
The linear-response contribution to the volume energy (per macroion)
associated with macroion-counterion interactions [Eq.~(\ref{E01})]
has a closely related form
\begin{equation}
v^{(2)}_{\rm ind}(0)=\int{\rm d}{\bf r}_1\,\int{\rm d}{\bf r}_2\,
\chi(|{\bf r}_1-{\bf r}_2|)v_{mc}(r_1)v_{mc}(r_2),
\label{Er}
\end{equation}
and a similar physical interpretation, except that the induced density 
now interacts back with the first macroion, generating a one-body energy.

Proceeding from linear to nonlinear response, the effective triplet 
interaction can be expressed [from Eq.~(\ref{v3effk})] as
\begin{equation}
v^{(3)}_{\rm eff}({\bf r},{\bf r}')=2\int{\rm d}{\bf r}_1\,
\int{\rm d}{\bf r}_2\,\int{\rm d}{\bf r}_3\,
\chi'({\bf r}_1-{\bf r}_3,{\bf r}_2-{\bf r}_3)
v_{mc}(r_1)v_{mc}(|{\bf r_2}-{\bf r}|)v_{mc}(|{\bf r_3}-{\bf r}'|).
\label{v3effr-1}
\end{equation}
Again the interpretation is clear: the external potentials due to two macroions
(top two macroions in Fig.~\ref{fig-diagram}), separated by displacement 
${\bf r}$, induce a change in counterion density at point ${\bf r}_3$.  
The induced density, which depends via $\chi'$ on triplet density 
correlations in the OCP, then interacts with a third macroion, at displacement 
${\bf r}'$ from the first, contributing a counterion-induced three-particle 
interaction energy.
Considering now the nonlinear correction to the pair interaction, 
and leaving aside the term arising from charge neutrality, the main 
contribution can be written [from Eq.~(\ref{Delta-v2effk})] as
\begin{equation}
\Delta v^{(2)}_{\rm eff}(r)=2\int{\rm d}{\bf r}_1\,\int{\rm d}{\bf r}_2\,
\int{\rm d}{\bf r}_3\,\chi'({\bf r}_1-{\bf r}_3,{\bf r}_2-{\bf r}_3)
v_{mc}(r_1)v_{mc}(r_2)v_{mc}(|{\bf r_3}-{\bf r}|).
\label{Delta-v2effr-1}
\end{equation}
The interpretation is analogous to that for the triplet interaction,
except that the external potentials at points ${\bf r}_1$ and ${\bf r}_2$ 
are now associated with the same macroion.
Finally, the nonlinear correction to the volume energy [Eq.~(\ref{Delta-E1})], 
aside from the charge neutrality term, has the form
\begin{equation}
\Delta E=\frac{N_m}{3}\int{\rm d}{\bf r}_1\,\int{\rm d}{\bf r}_2\,
\int{\rm d}{\bf r}_3\,\chi'({\bf r}_1-{\bf r}_3,{\bf r}_2-{\bf r}_3)
v_{mc}(r_1)v_{mc}(r_2)v_{mc}(r_3).
\label{Delta-Er}
\end{equation}
The physical meaning of $\Delta E$ is similar to that of 
$\Delta v^{(2)}_{\rm eff}(r)$, except that now the density that is 
induced nonlinearly by one macroion interacts back with the same macroion,
generating a nonlinear contribution to the one-body energy.

\subsection{Random Phase Approximation}\label{RPA}
Further progress towards practical expressions for effective interactions
requires specifying the OCP response functions.
For charged colloids, the OCP is typically weakly correlated, 
characterized by relatively small coupling parameters:
$\Gamma=\lambda_B/a_c \ll 1$, where
$\lambda_B=\beta z^2e^2/\epsilon$ is the Bjerrum length and
$a_c=(3/4\pi n_c)^{1/3}$ is the counterion-sphere radius.
For example, for macroions of valence $Z=500$, volume fraction $\eta=0.01$,
and monovalent counterions suspended in salt-free water at room temperature 
($\lambda_B=0.714$ nm), we find $\Gamma\simeq 0.02$.  
For such weakly-correlated plasmas, 
it is reasonable -- at least as regards long-range interactions -- 
to neglect short-range correlations.  We thus adopt the random phase
approximation (RPA), which equates the two-particle direct
correlation function (DCF) to its exact asymptotic limit:
$c^{(2)}(r)=-\beta v_{cc}(r)$ or
$\hat c^{(2)}(k)=-4\pi\beta z^2 e^2/\epsilon k^2$. 
In neglecting short-range correlations, the RPA is formally equivalent 
to the mean-field PB theory, as shown in the Appendix.
Furthermore, we ignore the influence of the macroion hard cores on
the OCP response functions, which is reasonable for
sufficiently dilute suspensions.
Within the RPA, the OCP (two-particle) static structure factor 
and linear response function take the analytical forms
\begin{equation}
S(k)=\frac{1}{1-n_c \hat c^{(2)}(k)}=\frac{1}{1+\kappa^2/k^2}
\label{Sk}
\end{equation}
and
\begin{equation}
\chi(k)=-\beta n_cS(k)=\frac{-\beta n_c}{1+\kappa^2/k^2}, \label{chi1}
\end{equation}
where $\kappa=\sqrt{4\pi n_cz^2e^2/\epsilon k_BT}$.
As will be seen below, the parameter $\kappa$ plays the role of
the Debye screening constant (inverse screening length) in
the counterion density profile and in the effective interactions.
In the absence of salt, the counterions are the only screening ions.
The macroions themselves, being singled out as sources of the 
external potential for the counterions, do not contribute to the 
density of screening ions.  
Fourier transforming Eq.~(\ref{chi1}), the real-space linear response 
function takes the form
\begin{equation}
\chi(r)=-\beta n_c\left[\delta({\bf r})+n_c h_{cc}(r)\right],
\label{chir}
\end{equation}
where
\begin{equation}
h_{cc}(r)=-\frac{\beta z^2e^2}{\epsilon} \frac{e^{-\kappa r}}{r}
\label{hr}
\end{equation}
is the counterion-counterion pair correlation function~\cite{note2},
which has Yukawa form, with screening constant $\kappa$.
Equation~(\ref{chir}) makes clear that there are two physically 
distinct types of counterion response: local response, associated with 
counterion self correlations, and nonlocal response, associated with 
counterion pair correlations.

At this point, we can specify the constant $\rho_0$ in Eq.~(\ref{delta-rho}).  
Combining Eq.~(\ref{chi1}) with the long-wavelength limit of the 
macroion-counterion interaction, $\hat v_{mc}(k)\to 4\pi Zze^2/\epsilon k^2$, 
as $k\to 0$, we have
\begin{equation}
\lim_{k\to 0}\left[\chi(k)\hat v_{mc}(k)\right]=Z/z.
\end{equation}
Thus, the linear response term in Eq.~(\ref{delta-rho}) already ensures 
proper normalization of $\rho_c({\bf r})$, which implies that $\rho_0=0$.

Proceeding to nonlinear response, we first note that
the three-particle structure factor obeys the identity
\begin{equation}
S^{(3)}({\bf k},{\bf k}')=S(k)S(k')S(|{\bf k}+{\bf k}'|)
\left[1+n_c^2~\hat c^{(3)}({\bf k},{\bf k}')\right],
\label{S3}
\end{equation}
where $\hat c^{(3)}({\bf k},{\bf k}')$ is the Fourier transform of
the three-particle direct correlation function.  Within the RPA,
however, $c^{(3)}$ and all higher-order DCF's vanish.
Thus, from Eqs.~(\ref{chi1k}), (\ref{chi2k}), and (\ref{S3}), the first 
nonlinear response function can be expressed in Fourier space as
\begin{equation}
\chi'({\bf k},{\bf k}')=-\frac{k_{\rm B}T}{2n_c^2}\chi(k)\chi(k')
\chi(|{\bf k}+{\bf k}'|)
\label{chi2}
\end{equation}
and in real space as
\begin{equation}
\chi'({\bf r}_1-{\bf r}_2,{\bf r}_1-{\bf r}_3)=-\frac{k_{\rm B}T}{2n_c^2}
\int{\rm d}{\bf r}\,\chi(|{\bf r}_1-{\bf r}|)\chi(|{\bf r}_2-{\bf r}|)
\chi(|{\bf r}_3-{\bf r}|).
\label{chi2r}
\end{equation}
Higher-order nonlinear counterion response leads to higher-order terms 
in the effective Hamiltonian [Eq.~(\ref{Heff2})].
For example, the effective four-body interaction takes the form
\begin{equation}
v_{\rm eff}^{(4)}({\bf k},{\bf k}',{\bf k}'')
=6\chi''({\bf k}',{\bf k}'',-{\bf k}-{\bf k}'-{\bf k}'') 
\hat v_{mc}(k)\hat v_{mc}(k')
\hat v_{mc}(k'') \hat v_{mc}(|{\bf k}+{\bf k}'+{\bf k}''|),
\label{v4effk}
\end{equation}
where
\begin{equation}
\chi''({\bf k},{\bf k}',{\bf k}'')=\frac{-\beta^3}{3!}n_c
S^{(4)} ({\bf k},{\bf k}',{\bf k}'') \label{chi3}
\end{equation}
is the next higher-order nonlinear response function and
\begin{eqnarray}
S^{(4)}({\bf k},{\bf k}',{\bf k}'')~&=&~S(k)S(k')S(k'')S(|{\bf k}+
{\bf k}'+{\bf k}''|) \nonumber \\
~&\times&~\left[S(|{\bf k}+{\bf k}'|)+S(|{\bf k}'+{\bf
k}''|)+S(|{\bf k}+{\bf k}''|)-2\right] \label{S4}
\end{eqnarray}
is the four-particle structure factor in the RPA.
Just as effective three-body interactions are related to corrections at 
the two- and one-body levels, so four-body interactions entail corrections 
at the three-, two-, and one-body levels, which (in Fourier space) 
are proportional to appropriate summations of
$v_{\rm eff}^{(4)}({\bf k},{\bf k}',{\bf k}'')$ over the 
wave-vectors ${\bf k}$, ${\bf k}'$, and ${\bf k}''$.  
In principle, these higher-order corrections could be computed 
to further check for convergence of the perturbation expansion.

\section{Results}\label{Results}

\subsection{Counterion Density}\label{Counterion Density}

A practical expression for the ensemble-averaged counterion density is now 
obtained by substituting the linear and first nonlinear RPA response functions 
[Eqs.~(\ref{chi1}) and (\ref{chi2})] into the expansion 
for $\la\hat\rho_c({\bf k})\ra$ [Eq.~(\ref{rhock2})].  
The result may be expressed in the form
\begin{equation}
\la\hat\rho_c({\bf k})\ra=\hat\rho_{c0}({\bf k})-\frac{\chi(k)}
{2\beta n_c^2 V'}\sum_{{\bf k}'} \hat\rho_{c0}({\bf k}') 
\hat\rho_{c0}({\bf k}-{\bf k}'), \qquad k\neq 0, \label{rhock3}
\end{equation}
where
\begin{equation}
\hat\rho_{c0}({\bf k})=\chi(k)\hat
v_{mc}(k)\hat\rho_m({\bf k}), \qquad k\neq 0, \label{rhoc1k}
\end{equation}
is the Fourier transform of the linear-response counterion density 
and $\hat v_{mc}(k)$ is the transform of the macroion-counterion interaction 
(specified below).  Inverse transforming Eq.~(\ref{rhock3}) yields
\begin{equation}
\la\rho_c({\bf r})\ra=\rho_{c0}({\bf r})-\frac{1}{2\beta
n_c^2}\int{\rm d}{\bf r}'\,\chi(|{\bf r}-{\bf
r}'|)[\rho_{c0}({\bf r}')]^2, \label{rhocr1}
\end{equation}
where 
\begin{equation}
\rho_{c0}({\bf r})=\sum_{i=1}^{N_m}\rho_1(|{\bf r}-{\bf R}_i|)
\label{rhocrlin1}
\end{equation}
is the real-space linear response counterion density in the presence of 
macroions fixed at positions ${\bf R}_i$, expressed as a sum of 
single-macroion counterion density orbitals $\rho_1(r)$ --
the inverse transform of $\hat\rho_1(k)=\chi(k)\hat v_{mc}(k)$.
Equivalently, 
\begin{equation}
\rho_{c0}({\bf r})=\int{\rm d}{\bf r}'\,\chi(|{\bf r}-{\bf r}'|)
\sum_{i=1}^{N_m} v_{mc}(|{\bf r}'-{\bf R}_i|).
\label{rhocrlin2}
\end{equation}
Now substitution of Eqs.~(\ref{chir}) and (\ref{hr}) for the real-space RPA 
linear response function into Eqs.~(\ref{rhocr1}) and (\ref{rhocrlin2}) 
allows the linear-response counterion density profile to be expressed as
\begin{equation}
\rho_{c0}({\bf r})=\beta n_c\sum_{i=1}^{N_m}\left[
-v_{mc}(|{\bf r}-{\bf R}_i|)+\frac{\kappa^2}{4\pi} \int {\rm d}{\bf r}'\, 
\frac{e^{-\kappa|{\bf r}-{\bf r}'|}}{|{\bf r}-{\bf r}'|}
v_{mc}(|{\bf r}'-{\bf R}_i|)\right]
\label{rhocr-explicit}
\end{equation}
and the nonlinear profile as
\begin{equation}
\la\rho_c({\bf r})\ra=\rho_{c0}({\bf r})+\frac{1}{2n_c}
\left[\rho_{c0}({\bf r})\right]^2-\frac{\kappa^2}{8\pi n_c}
\int{\rm d}{\bf r}'\,\frac{e^{-\kappa |{\bf r}-{\bf r}'|}}{|{\bf r}-{\bf r}'|}
\left[\rho_{c0}({\bf r}')\right]^2. \label{rhocr2}
\end{equation}
The last two terms on the right side of Eq.~(\ref{rhocr2}) are nonlinear 
corrections to the linear profile and can be physically interpreted 
as arising, respectively, from local and nonlocal nonlinear response of 
counterions to the macroion charge. 

The linear-response counterion profile in the presence of a 
{\it distribution} of macroions can be obtained by ensemble averaging 
Eq.~(\ref{rhoc1k}) over macroion coordinates.  In Fourier space,
\begin{equation}
\la\hat\rho_{c0}({\bf k})\ra_m=\rho_1(k)\la\hat\rho({\bf k})\ra_m,
\label{rhockav}
\end{equation}
where $\la~\ra_m$ again represents a trace over macroion coordinates and 
$\la\hat\rho({\bf k})\ra_m$ is the Fourier component of the average density 
of macroions.  In real space, the average density of counterions around 
a central macroion can then be expressed as
\begin{equation}
\la\rho_{c0}({\bf r})\ra_m=\rho_1({\bf r})+n_m\int{\rm d}{\bf r}'\,
g_{mm}({\bf r}')\rho_1(|{\bf r}-{\bf r}'|),
\label{rhocrav}
\end{equation}
where $g_{mm}({\bf r})$ is the macroion-macroion pair distribution function.
The latter function may be obtained from integral-equation theory or
simulation, with effective interactions as input.

\subsection{Macroion-Counterion Interaction}\label{mcint}
To this point, the theory makes no assumptions about the type of macroion.  
Practical calculations require specifying the macroion structure, 
the macroion-counterion interaction, and the corresponding single-macroion 
counterion density orbital.  Henceforth, we specialize to 
charged hard-sphere colloidal macroions.  A convenient strategy, 
proposed in ref.~\cite{vRH} and adopted in refs.~\cite{Denton1} and
\cite{Denton2}, specifies $v_{mc}(r)$ inside the hard core ($r<a$) 
so as to minimize counterion penetration.  We thus assume
\begin{equation}
v_{mc}(r)=\left\{ \begin{array} {l@{\quad\quad}l}
\frac{\displaystyle -Zze^2}{\displaystyle \epsilon r}, & r>a \\
\frac{\displaystyle -Zze^2}{\displaystyle \epsilon a}\alpha, & r<a
\end{array} \right.
\label{vmcr}
\end{equation}
and choose the parameter $\alpha$ appropriately.
In passing, we note that $v_{mc}(r)$ plays a role here analogous to 
that of an empty-core pseudopotential in the pseudopotential theory 
of simple metals~\cite{Hafner87,Ashcroft66}.
As shown in refs.~\cite{vRH} and \cite{Denton1}, at the level of 
linear response, penetration of counterions inside the macroion cores 
is eliminated by choosing $\alpha=\kappa a/(1+\kappa a)$.  
This choice yields
\begin{equation}
\hat v_{\rm mc}(k)=-\frac{4\pi Zze^2}{\epsilon(1+\kappa a)k^2}
\left[\cos(ka)+\frac{\kappa}{k} \sin(ka) \right]
\label{vmck2}
\end{equation}
and
\begin{equation}
\rho_1(r)=\left\{ \begin{array}
{l@{\quad\quad}l}
\frac{\displaystyle Z}{\displaystyle z}\frac{\displaystyle \kappa^2}
{\displaystyle 4\pi}
~\frac{\displaystyle e^{\kappa a}}{\displaystyle 1+\kappa a}
~\frac{\displaystyle e^{-\kappa r}}{\displaystyle r}, & r>a \\
0, & r<a,
\end{array} \right.
\label{rho1r2}
\end{equation}
which is precisely the DLVO expression for the density of counterions 
around an isolated macroion~\cite{DL,VO}. 
The above choice for the parameter $\alpha$ ensures that the linear
term and first nonlinear term of Eq.~(\ref{rhocr2}) vanish completely
inside the macroion core.  The same parametrization also allows, however, 
the final nonlinear term in Eq.~(\ref{rhocr2}) to be nonzero inside the core, 
although in practice the fractional penetration is at most a few percent.  
Independent of parametrization, Eqs.~(\ref{rhocr1}), (\ref{rhocrlin1}), 
and (\ref{rhocr2}) maintain charge neutrality by preserving the number 
of counterions, since $\int{\rm d}{\bf r}\,\rho_1({\bf r})=Z/z$ and 
$\int{\rm d}{\bf r}\, \chi(r)=\chi(k=0)=0$.

Another artifact of the present scheme, apparent from Eq.~(\ref{rhocrlin1}), 
is that the counterion density profile around a given macroion overlaps 
the hard cores of neighboring macroions.  More general parametrizations of 
the macroion-counterion interaction than Eq.~(\ref{vmcr}) could conceivably 
eliminate counterion penetration within all cores.  
An alternative strategy would incorporate excluded volume constraints 
directly into the response functions, 
which then would more properly describe the Swiss cheese OCP. 
In such a scheme, $\chi(|{\bf r}-{\bf r}'|)$ would strictly 
vanish when either ${\bf r}$ or ${\bf r}'$ falls inside a hard core.  
This condition -- not obeyed by Eqs.~(\ref{chi1}) and (\ref{chi2}) -- 
would enforce exclusion of counterions from all macroion cores.
Nevertheless, in the current scheme, the extent of core overlap 
is minor for macroion separations that significantly exceed the screening 
length $\kappa^{-1}$, which is usually the case in practice.

\subsection{Effective Interactions}\label{Effective Interactions}
The effective interactions can be expressed in real space
by evaluating the respective inverse Fourier transforms.
From Eqs.~(\ref{v3effk}) and (\ref{chi2}), the effective triplet 
interaction is
\begin{equation}
v^{(3)}_{\rm eff}({\bf r}_1-{\bf r}_2,{\bf r}_1-{\bf r}_3)
=-\frac{k_BT}{n_c^2}\int{\rm d}{\bf r}\,\rho_1(|{\bf r}_1-{\bf
r}|)\rho_1(|{\bf r}_2-{\bf r}|)\rho_1(|{\bf r}_3-{\bf r}|).
\label{v3effr-2}
\end{equation}
Equations~(\ref{v2indk}) and (\ref{v20k}), combined with 
Eqs.~(\ref{chi1}) and (\ref{vmck2}), yield the linear-response 
prediction for the effective pair interaction~\cite{Denton1},  
\begin{equation}
v^{(2)}_0(r)=\frac{Z^2e^2}{\epsilon}\left(\frac{e^{\kappa a}}
{1+\kappa a}\right)^2~\frac{e^{-\kappa r}}{r}, \quad r>\sigma, 
\label{v20r}
\end{equation}
identical to the familiar DLVO screened-Coulomb potential in the limit of
widely separated macroions~\cite{DL,VO}, 
while Eq.~(\ref{Delta-v2effk}) yields the first nonlinear correction 
\begin{equation}
\Delta v^{(2)}_{\rm eff}(r)=-\frac{k_BT}{n_c^2}\int{\rm d}{\bf
r}'\,\rho_1(r')\rho_1(|{\bf r}-{\bf r}'|)\left[\rho_1(|{\bf
r}-{\bf r}'|)-\frac{n_c}{3}\right]. \label{Delta-v2effr-2}
\end{equation}
The total effective pair potential is given by
$v^{(2)}_{\rm eff}(r)=v^{(2)}_0(r)+\Delta v^{(2)}_{\rm eff}(r)$.
Note the distinction between the effective pair potential, which is
the interaction between a pair of macroions in a colloidal suspension 
of arbitrary concentration, and the potential of mean force,  
which is the interaction between an isolated pair of macroions, 
{\it i.e.}, the low-density limit of $v^{(2)}_{\rm eff}(r)$.

The volume energy can be expressed -- by combining 
Eqs.~(\ref{v2indk}), (\ref{v3effk}), (\ref{E01}), (\ref{Delta-E1}), and 
(\ref{vmck2}) -- as the sum of the linear response prediction~\cite{Denton1},
\begin{equation}
E_0=F_{\rm OCP}-N_m\frac{Z^2e^2}{2\epsilon}
\frac{\kappa}{1+\kappa a}-\frac{N_ck_BT}{2}, \label{E02}
\end{equation}
and the first nonlinear correction,
\begin{equation}
\Delta E=-\frac{N_mk_BT}{6n_c^2}\left(
\int{\rm d}{\bf r}\,\left[\rho_1(r)\right]^3
-n_c\int{\rm d}{\bf r}\,\left[\rho_1(r)\right]^2\right).
\label{Delta-E1r}
\end{equation}
The first and second terms on the right side of Eq.~(\ref{E02})
account, respectively, for the counterion entropy and the
macroion-counterion electrostatic interaction energy, while
Eq.~(\ref{Delta-E1r}) corrects the latter term for nonlinear response.
The final terms on the right sides of 
Eqs.~(\ref{Delta-v2effr-2})-(\ref{Delta-E1r})
originate from the charge neutrality constraint.

\subsection{Effect of Added Salt}
Finally, we generalize the above results to the case of nonzero salt 
concentration.  The average number density (in the free volume) of 
salt ion pairs, $n_s=N_s/V'$, is supposed maintained by exchange of 
salt ions with a salt reservoir through a semi-permeable membrane.
The total average microion density is then $n_{\mu}=n_++n_-=n_c+2n_s$, 
where $n_{\pm}$ are the average number densities of positive/negative microions.
Following ref.~\cite{Denton2}, the Hamiltonian generalizes to
\begin{equation}
H=H_m+H_{\mu}+H_{m+}+H_{m-},
\label{Hsalt}
\end{equation}
where $H_{\mu}$ is the Hamiltonian of the microions (counterions plus
salt ions) and $H_{m\pm}$ are the electrostatic interaction energies 
between macroions and positive/negative microions.
The presence of positive and negative microion species requires
a proliferation of response functions, $\chi_{ij}$ and $\chi'_{ijk}$, 
$i,j,k=\pm$, and a generalization of Eq.~(\ref{chir}) to
\begin{equation}
\chi_{++}(r)=-\beta n_+\left[\delta({\bf r})+n_+ h_{++}(r)\right],
\label{chi++r}
\end{equation}
\begin{equation}
\chi_{+-}(r)=-\beta n_+n_- h_{+-}(r),
\label{chi+-r}
\end{equation}
\begin{equation}
\chi_{--}(r)=-\beta n_-\left[\delta({\bf r})+n_- h_{--}(r)\right],
\label{chi--r}
\end{equation}
where $h_{ij}(r)$, $i,j=\pm$, are the bulk microion two-particle 
pair correlation functions, which depend implicitly on $n_{\pm}$.  
Generalizing Eq.~(\ref{rhock2}), the ensemble-averaged microion 
number densities are given by
\begin{eqnarray}
\la\hat\rho_{\pm}({\bf k})\ra~&=&~\pm\chi_{\pm}(k)
\hat v_{m+}(k)\hat\rho_m({\bf k})+\frac{1}{V'}\sum_{{\bf k}'}
\chi'_{\pm}({\bf k}',{\bf k}-{\bf k}')
\hat v_{m+}(k') \hat v_{m+}(|{\bf k}-{\bf k}'|) \nonumber \\
~&\times&~\hat\rho_m({\bf k}') \hat\rho_m({\bf k}-{\bf k}')
+\cdots \qquad (k\neq 0), \label{rho+-k}
\end{eqnarray}
where we have exploited symmetries:
$\hat v_{m+}=-\hat v_{m-}$, 
$\chi_{+-}=\chi_{-+}$, $\chi'_{++-}=\chi'_{+-+}$, 
etc., to define composite response functions as 
$\chi_+=\chi_{++}-\chi_{+-}$, $\chi_-=\chi_{--}-\chi_{+-}$, 
$\chi'_+=\chi'_{+++}-2\chi'_{++-}+\chi'_{+--}$, and 
$\chi'_-=\chi'_{---}-2\chi'_{-+-}+\chi'_{-++}$.
Substituting Eq.~(\ref{rho+-k}) into the multi-component Hamiltonian
[Eq.~(\ref{Hsalt})], the macroion-microion interaction contribution 
can be expressed as
\begin{eqnarray}
H_{m+}+H_{m-}~&=&~
\frac{1}{V'}\sum_{{\bf k}}\chi(k)\left[\hat v_{m+}(k)\right]^2
\hat\rho_m({\bf k})\hat\rho_m(-{\bf k}) 
+\frac{1}{V'^2}\sum_{{\bf k},{\bf k}'}\chi'({\bf k}',-{\bf k}-{\bf k}')
\nonumber \\
~&\times&~\hat v_{m+}(k)\hat v_{m+}(k')\hat v_{m+}(|{\bf k}+{\bf k}'|) 
\hat\rho_m({\bf k})\hat\rho_m({\bf k}')\hat\rho_m(-{\bf k}-{\bf k}'),
\label{H+-}
\end{eqnarray}
where the linear and first nonlinear response functions are now redefined 
as the following combinations of $\chi_{ij}$ and $\chi'_{ijk}$: 
\begin{equation}
\chi=\chi_+ +\chi_-=\chi_{++}-2\chi_{+-}+\chi_{--}
\end{equation}
and
\begin{equation}
\chi'=\chi'_+ -\chi'_-=\chi'_{+++}-3\chi'_{++-}+3\chi'_{+--}-\chi'_{---}.
\label{chip}
\end{equation}
The net effect of adding salt is to modify the previous salt-free results 
as follows.
First, the average counterion density $n_c$ in the Debye screening constant 
$\kappa$ and in the linear response function [Eq.~(\ref{chi1})] must be
replaced by the total average microion density $n_{\mu}$.
Thus, $\kappa\to \sqrt{4\pi n_{\mu}z^2e^2/\epsilon k_BT}$ and
$\chi(k)\to -\beta n_{\mu}S(k)$.
The first nonlinear response function retains its original form
[Eq.~(\ref{chi2})], but with the new definition of $\kappa$.
The second modification is in the linear-response volume energy 
[Eq.~(\ref{E02})], which becomes~\cite{Denton2,note3}
\begin{equation}
E_0=F_{\rm plasma}-N_m\frac{Z^2e^2}{2\epsilon}
\frac{\kappa}{1+\kappa a}-\frac{k_BT}{2}\frac{(N_+-N_-)^2}{N_++N_-}, 
\label{E03}
\end{equation}
where $F_{\rm plasma}$ is the free energy of the unperturbed microion plasma.
Finally, the effective triplet interaction and nonlinear
corrections to the effective pair interaction and volume energy
are generalized as follows:
\begin{equation}
\Delta E=-\frac{N_mk_BT}{6}\frac{(n_+-n_-)}{n_{\mu}^3}\left(
\int{\rm d}{\bf r}\,\left[\rho_1(r)\right]^3
-n_{\mu}\int{\rm d}{\bf r}\,\left[\rho_1(r)\right]^2\right),
\label{Delta-E1r-salt}
\end{equation}
\begin{equation}
\Delta v^{(2)}_{\rm eff}(r)=-k_BT\frac{(n_+-n_-)}{n_{\mu}^3}\int{\rm d}
{\bf r}'\,\rho_1(r')\rho_1(|{\bf r}-{\bf r}'|)\left[\rho_1(|{\bf r}
-{\bf r}'|)-\frac{n_{\mu}}{3}\right],
\label{Delta-v2effr-salt}
\end{equation}
\begin{equation}
v^{(3)}_{\rm eff}({\bf r}_1-{\bf r}_2,{\bf r}_1-{\bf r}_3)
=-k_BT\frac{(n_+-n_-)}{n_{\mu}^3}\int{\rm d}{\bf r}\,
\rho_1(|{\bf r}_1-{\bf r}|)\rho_1(|{\bf r}_2-{\bf r}|)
\rho_1(|{\bf r}_3-{\bf r}|). \label{v3effr-salt}
\end{equation}

Equations~(\ref{Delta-E1r-salt})-(\ref{v3effr-salt}) 
[combined with Eq.~(\ref{rho1r2}) for $\rho_1(r)$]
are the main new results for nonlinear effective interactions.
These expressions imply that nonlinear effects increase in strength 
with increasing macroion charge, increasing macroion concentration, 
and decreasing salt concentration, and that effective triplet interactions 
are consistently attractive.  
These results also imply that in the limit of zero macroion concentration 
($n_c=n_+-n_-\to 0$), or of high salt concentration ($n_{\mu}\to\infty$), 
such that $(n_+-n_-)/n_{\mu}\to 0$, the leading-order nonlinear corrections 
all vanish.  This dilute limit follows naturally from the fact that, 
for a pure symmetric electrolyte, response functions related by symmetry 
are equal ($\chi'_{+++}=\chi'_{---}$, $\chi'_{++-}=\chi'_{+--}$), and so,
from Eq.~(\ref{chip}), the nonlinear response function $\chi'$ is zero.
This result -- a consequence of charge neutrality that is analogous to 
the vanishing of the first nonlinear (quadratic) term in the expansion 
of the nonlinear PB equation -- may partially explain the surprisingly 
broad range of validity of DLVO theory for high-ionic-strength suspensions.
Nevertheless, even in dilute suspensions at high ionic strength, 
higher-order nonlinear corrections do not necessarily vanish.  
For this reason, predictions of the first-order nonlinear theory 
in the dilute limit may deviate from numerical solutions of 
the full nonlinear PB equation, which include, by construction, nonlinear 
corrections to all orders.  The first-order corrections are nonetheless 
valuable in signaling the onset of nonlinearity, as shown below.

Equations (\ref{Delta-v2effr-salt}) and (\ref{v3effr-salt}) may be compared 
with related expressions derived via the density-functional approach.  
Equation~(\ref{Delta-v2effr-salt}) is similar in structure to an expression 
for a wall-induced effective pair interaction derived by Goulding and 
Hansen~\cite{Goulding99,HGvR00} (Eq.~(13) of Ref.~\cite{Goulding99})
if one factor of $\rho_1(r)$ in Eq.~(\ref{Delta-v2effr-salt})
is replaced by a wall counterion density orbital.
It should be noted that these authors neglected the bulk nonlinear 
correction derived here [Eq.~(\ref{Delta-v2effr-salt})], which is 
justified only in the dilute limit of an isolated pair of macroions.
Equation~(\ref{v3effr-salt}) is similar to an expression 
for the triplet interaction derived by L\"owen and Allahyarov~\cite{LA}, 
differing only by a factor of $(n_+-n_-)/n_{\mu}$ and by our 
excluded-volume correction in the definition of $\kappa$.

\subsection{Analytical Expressions for Effective Interactions}
Quantitative predictions of nonlinear response theory are facilitated 
by reducing the effective interactions to computationally practical 
analytical forms.  Substituting Eq.~(\ref{rho1r2}) into 
Eq.~(\ref{Delta-E1r-salt}), and evaluating the integrals, the 
nonlinear correction to the volume energy can be expressed as
\begin{equation}
\Delta E=\frac{N_mk_BT}{6}\frac{(n_+-n_-)}{n_{\mu}^3}\left[
\frac{Z^2\kappa^3n_{\mu}}{8\pi}\left(\frac{1}{1+\kappa a}\right)^2
-\frac{Z^3\kappa^6}{(4\pi)^2}
\left(\frac{e^{\kappa a}}{1+\kappa a}\right)^3
{\rm E}_1(3\kappa a)\right],
\label{Delta-E1r-analytical}
\end{equation}
where ${\rm E}_1$ is the exponential integral function~\cite{GR}
\begin{equation}
{\rm E}_1(x)=\int_1^{\infty}{\rm d}u\,\frac{e^{-xu}}{u}, \qquad x>0.
\label{E1x}
\end{equation}
Similarly, the nonlinear correction to the effective pair potential 
can be rendered analytically.  The key is expressing the integral
$I_1(r)\equiv\int{\rm d}{\bf r}'\rho_1(r')[\rho_1(|{\bf r}-{\bf r}'|)]^2$
in Eq.~(\ref{Delta-v2effr-salt}) in the form 
$I_1={\cal F}^{-1}\{\hat\rho_1(k)\hat\rho_2(k)\}$, where
$\hat\rho_2(k)\equiv{\cal F}\{[\rho_1(r)]^2\}$, with ${\cal F}$ denoting
the Fourier transform operator and ${\cal F}^{-1}$ its inverse.
Substituting Eq.~(\ref{rho1r2}) into Eq.~(\ref{Delta-v2effr-salt}), 
$\Delta v_{\rm eff}^{(2)}(r)$ reduces, after some algebra, to
\begin{equation}
\Delta v^{(2)}_{\rm eff}(r)=f_1(r)~\frac{e^{-\kappa r}}{r}
+f_2(r)~\frac{e^{\kappa r}}{r}+f_3(r)~\frac{e^{-\kappa a}}{r}, \quad r>\sigma,
\label{Delta-v2effr-analytical}
\end{equation}
where
\begin{equation}
f_1(r)=C_1\left[\kappa(r-\sigma)+1-e^{-\kappa\sigma}\right]
+C_2\left[{\rm E}_1\left(\kappa(r-a)\right)
+{\rm E}_1\left(3\kappa a\right)
-{\rm E}_1\left(\kappa a\right)\right],
\end{equation}
\begin{equation}
f_2(r)=-C_2~{\rm E}_1\left(3\kappa(r+a)\right),
\end{equation}
and
\begin{equation}
f_3(r)=C_2~\left[{\rm E}_1(2\kappa(r+a))-{\rm E}_1(2\kappa(r-a))\right],
\label{f3}
\end{equation}
with
\begin{equation}
C_1=\frac{1}{6}\frac{(n_+-n_-)}{n_{\mu}}\frac{Z^2e^2}{\epsilon}\left(
\frac{e^{\kappa a}}{1+\kappa a}\right)^2
\label{C1}
\end{equation}
and
\begin{equation}
C_2=\frac{1}{8\pi}\frac{(n_+-n_-)}{n_{\mu}^2}\frac{Z^3e^2\kappa^3}{z\epsilon}
\left(\frac{e^{\kappa a}}{1+\kappa a}\right)^3.
\label{C2}
\end{equation}

It is interesting to examine the asymptotic ($r\to\infty$) behavior of 
the leading-order nonlinear response approximation to the effective pair 
interaction.  From Eqs.~(\ref{Delta-v2effr-analytical})-(\ref{f3}), 
using the inequality ${\rm E}_1(x)<e^{-x}/x$, we find the asymptotic limit 
\begin{equation}
\lim_{r\to\infty}v^{(2)}_{\rm eff}(r)=C_1\kappa~e^{-\kappa r},
\label{v2effr-asymptotic}
\end{equation}
which exhibits a more gradual decay than the screened-Coulomb DLVO potential 
[Eq.~(\ref{v20r})].
We emphasize that this result does not contradict measurements of DLVO-like
interactions between isolated pairs of macroions in dilute 
suspensions~\cite{Grier1}, since in the dilute limit $C_1\to 0$ and
the asymptotic behavior reduces to that of linear response. 
The physical significance of Eq.~(\ref{v2effr-asymptotic}) may be limited, 
however, by the neglect of higher-order nonlinear terms and shielding 
by intervening macroions~\cite{Klein02} at distances beyond the mean 
nearest-neighbor separation, $r>[3/(4\pi n_m)]^{1/3}$.

\section{Numerical Investigations and Discussion}\label{Numerical}
To quantitatively illustrate the influence of nonlinear screening,
we compute counterion density profiles and effective pair and triplet 
interactions for selected system parameters.
All results presented are for the case of monovalent counterions ($z=1$)
and aqueous suspensions at room temperature ($\lambda_B=0.714$ nm).
Figure~\ref{fig-rhocr} compares the counterion density profile around a
single macroion of diameter $\sigma=100$ nm and valences $Z=100$ and 500
in a dilute ($\eta=0.01$) salt-free suspension, as predicted by linear 
response (DLVO) theory [Eq.~(\ref{rhocrlin2})] and by first-order nonlinear 
response theory [Eq.~(\ref{rhocr2})]. The linear-response counterion density
$\rho_{c0}({\bf r})$ is approximated here by a single orbital $\rho_1(r)$
[Eq.~(\ref{rho1r2})].
Evidently, nonlinear response sharpens the distribution of counterions 
around a macroion.

Figure~\ref{fig-v2r} compares the linear and nonlinear response predictions for 
the effective pair interaction, now for $Z=400$ (a) and 700 (b) and with 
small concentrations of added salt. 
As a check on the calculations, the interactions were computed both 
by Monte Carlo integration of Eq.~(\ref{Delta-v2effr-salt}) and from
the analytical expressions [Eqs.~(\ref{Delta-v2effr-analytical})-(\ref{C2})],
the results being identical to within numerical error.
Enhanced screening, resulting from the sharper nonlinear counterion profiles 
around macroions, has the general effect of weakening the pair interactions.
For a given macroion diameter, nonlinear corrections increase in magnitude 
with increasing macroion valence and concentration and with decreasing 
salt concentration.  Qualitatively, our predictions are consistent with 
the recent observations of Durand and Franck~\cite{Franck} of surprisingly 
short-ranged pair correlations in highly deionized colloidal suspensions.  
A quantitative comparison, however, would require computing the radial
distribution function $g(r)$ from our $v_{\rm eff}^{(2)}(r)$, by means of
integral-equation theory or simulation, or computing $v_{\rm eff}^{(2)}(r)$
from the experimental $g(r)$ data.  Note that the effective pair potential 
discussed here is distinct from the potential of mean force $v_{\rm mf}(r)$, 
which was obtained in ref.~\cite{Franck} from the experimentally measured 
$g(r)$ via the definition $v_{\rm mf}(r)=-k_BT\ln g(r)$. 

In Fig.~\ref{fig-v2r}c, the parameters (macroion diameter, $\sigma=652$ nm, 
and volume fraction, $\eta=0.0352$) are chosen to compare with 
the experiments of Larsen and Grier~\cite{Grier2} in which 
unusually long-lived metastable fcc crystallites were observed.  
The macroion valence here is set to the maximum consistent with 
charge renormalization~\cite{Alexander},
$Z^*\sim O(10)(a/\lambda_B)\simeq 5000$, 
assuming the charge of a macroion to be reduced in bulk compared with 
its value in isolation~\cite{note4}.
While the pair interaction remains repulsive, it is significantly weaker
than the DLVO prediction over a range comparable to the macroion diameter.
A weaker pair interaction could promote the influence of three-body 
attractions, as well as interactions ignored by mean-field theory, 
such as short-range counterion fluctuation-induced 
attractions~\cite{Holm,Shklovskii99}.
In this way, nonlinear response may contribute to explaining experimental 
evidence for apparent attractions between like-charged macroions.
The phase behavior of highly-charged colloidal crystals will be 
the subject of a future study.

The insets to Fig.~\ref{fig-v2r} illustrate the extent to which the 
effective pair interaction may be fit by a screened-Coulomb (DLVO) potential.
For sufficiently weak nonlinearity (Fig.~\ref{fig-v2r}a), 
$v^{(2)}_{\rm eff}(r)$ may be reasonably fit by a DLVO potential with 
the same screening constant but a lower (renormalized) effective charge.
The tendency of nonlinear screening to preserve the DLVO form of potential 
is consistent with conclusions from PB cell model 
calculations~\cite{Alexander}, {\it ab initio} simulations~\cite{Lowen,Tehver},
and optical tweezer experiments~\cite{Grier2}, all of which support the 
bulk DLVO potential in the weakly nonlinear regime.  In the more strongly 
nonlinear regime (Figs.~\ref{fig-v2r}b and \ref{fig-v2r}c), however, our 
calculations 
indicate that effective pair interactions may deviate significantly from DLVO 
form.  In this regime, if $v^{(2)}_{\rm eff}(r)$ can be fit at all by a DLVO 
potential, then it is only over a limited range and only by allowing 
renormalization of both the charge and screening constant.
Similar departures from DLVO behavior with increasing macroion-counterion
coupling strength have been predicted by integral-equation 
theories~\cite{Patey80,Belloni86,Khan87,Carbajal-Tinoco02,Outhwaite02,
Petris02,Anta}.

Although in the physically relevant range of macroion valence ($Z<Z^*$)
the predicted pair interaction is always purely repulsive, at higher 
(unphysical) valences ($Z>Z^*$) the leading-order nonlinear theory 
predicts that $v^{(2)}_{\rm eff}(r)$ can develop an attractive well 
at sufficiently high macroion concentrations.  
Mathematical proofs~\cite{Neu,Sader-Chan,Trizac} have recently shown,
however, that PB theory cannot yield pair attraction -- at least between 
a pair of isolated macroions.  Since the RPA used here is formally 
equivalent to mean-field PB theory (see Appendix), the emergence of an 
attractive pair potential is best interpreted as a sign that 
higher-order nonlinear terms must then be included. 

As a quantitative test of the nonlinear response theory, we compare 
predictions with available data from the {\it ab initio} simulations of 
Tehver {\it et al.}~\cite{Tehver}. 
By assuming a counterion density
orbital and ignoring counterion density fluctuations, the {\it ab
initio} approach provides the most direct test of the theory.
Figure~\ref{fig-epot}a presents the comparison for the total potential
energy of interaction between a pair of macroions, of diameter
$\sigma=106$ nm and valence $Z=200$, in a cubic box of length
530 nm with periodic boundary conditions (taking into account
image interactions) in the absence of salt.  The theory is in essentially 
perfect agreement with simulation, although nonlinear effects, 
for these parameters, are relatively weak.
Figure~\ref{fig-epot}b shows results for a higher valence ($Z=700$),
for which case simulation data are not yet available,
but where nonlinear effects are more prominent.
Further simulations of more highly charged macroions would more severely
test the theory -- in particular, convergence of the perturbation 
expansion -- in the strongly nonlinear regime.

To quantify the range of validity of the linear response approximation
and to measure the impact of nonlinear screening on thermodynamics, we 
calculate the magnitude of the leading-order nonlinear correction to the 
pair interaction $\Delta v^{(2)}_{\rm eff}(r)$ at the mean nearest-neighbor 
separation $r_{\rm nn}$, where pair interactions contribute most to the 
potential energy.  Figure~\ref{fig-vscan} maps out, in the space of 
macroion volume fraction $\eta$ and salt concentration $c_s$
(measured in $\mu$M), the boundary 
of the region within which $|\Delta v^{(2)}_{\rm eff}(r_{\rm nn})|$ 
exceeds typical thermal energies, for the fcc crystal structure: 
$r_{\rm nn}/\sigma=2^{-1/2}(2\pi/3\eta)^{1/3}$.  For points above the 
boundary curves, $|\Delta v^{(2)}_{\rm eff}(r_{\rm nn})| >1~k_BT$ 
(Fig.~\ref{fig-vscan}a) or $0.1~k_BT$ (Fig.~\ref{fig-vscan}b).
Points on the boundary curves in Fig.~\ref{fig-vscan} 
correspond to thermodynamic states for which a stable fcc crystal phase 
is predicted by simulations of model Yukawa systems~\cite{Robbins88}, 
although these simulations do not include influences of the volume energy.  
With increasing $Z$ and decreasing $c_s$, the threshold $\eta$ {\it decreases}.
Thus, nonlinear screening is anticipated to increasingly influence
thermodynamics with increasing macroion charge and concentration and with 
decreasing ionic strength -- just the conditions under which anomalous 
phase behavior has been observed~\cite{Tata,Ise,Matsuoka,Groehn00,Grier2}.

Moving beyond pair interactions, Fig.~\ref{fig-v3r} shows the effective
three-body interaction between a triplet of macroions arranged in
an equilateral triangle for $\sigma=100$ nm and two different valences,
$Z=500$ and 700.  The interactions were computed numerically by Monte Carlo 
integration of Eq.~({\ref{v3effr-salt}).
The strength of the interaction is seen to grow rapidly with increasing
macroion valence and with decreasing separation between macroion cores.
In a concentrated suspension of highly-charged macroions, effective 
many-body interactions may become significant.  In particular, 
as noted above, triplet attractions may well contribute to the 
surprising metastability of colloidal crystallites observed in 
deionized suspensions~\cite{Grier2}.

To again test the theory against simulation, we compute the force 
on a macroion in an equilateral-triangle configuration of 
three macroions as a function of the triangle edge length.  
On expanding a triangle from edge length $R-\Delta R/2$ to $R+\Delta R/2$, 
the energy changes by
\begin{equation}
\Delta U=3\left[v^{(2)}_{\rm eff}\left(R+\frac{\Delta R}{2}\right)
-v^{(2)}_{\rm eff}\left(R-\frac{\Delta R}{2}\right)\right]
+v^{(3)}_{\rm eff}\left(R+\frac{\Delta R}{2}\right)
-v^{(3)}_{\rm eff}\left(R-\frac{\Delta R}{2}\right).
\end{equation}
Since, as the triangle expands, each of the three macroions moves a distance 
$\Delta R/\sqrt{3}$ parallel to the total effective force $F$ acting on it, 
the change in energy also may be expressed as $\Delta U=-3F\Delta R/\sqrt{3}$.  
Equating the two expressions for $\Delta U$, the total force can be written as
\begin{equation}
F(R)=F^{(2)}(R)+F^{(3)}(R),
\label{Fr}
\end{equation}
where
\begin{equation}
F^{(2)}(R)=\frac{-\sqrt{3}}{\Delta R}\left[v^{(2)}_{\rm eff}
\left(R+\frac{\Delta R}{2}\right)
-v^{(2)}_{\rm eff}\left(R-\frac{\Delta R}{2}\right)\right]
\label{F2r}
\end{equation}
and
\begin{equation}
F^{(3)}(R)=\frac{-1}{\sqrt{3}\Delta R}\left[
v^{(3)}_{\rm eff}\left(R+\frac{\Delta R}{2}\right)
-v^{(3)}_{\rm eff}\left(R-\frac{\Delta R}{2}\right)\right]
\label{F3r}
\end{equation}
are the effective pair and triplet forces, respectively.

To compare directly with available simulation data~\cite{Tehver}, 
we consider macroions of diameter $\sigma=106$ nm in a cubic box 
of length 1000 nm with periodic boundary conditions in the absence
of salt.  Over a range of macroion valences, we compute the sum of 
linear (DLVO) effective pair forces, the sum of nonlinear effective 
pair forces, the effective triplet force, and the total effective 
force (sum of pair and triplet forces), from Eqs.~(\ref{Fr})-(\ref{F3r}).
For a valence of $Z=200$ -- the only case for which simulation data
were reported~\cite{Tehver} -- the predicted total force is 
essentially identical to the sum of pair forces, consistent with 
ref.~\cite{Tehver}, in which an absence of many-body effects was concluded.
For higher valences, however, three-body forces are {\it not} negligible.  
Figure~\ref{fig-f3r} presents predictions of the theory for $Z=700$ 
and 1000, demonstrating significant deviations of the total force 
from the sum of pair forces.  For the case $Z=1000$, which somewhat
exceeds the charge-renormalization limit~\cite{Alexander}, the predicted
total force actually becomes attractive beyond $r\simeq 2\sigma$, 
although this is likely an artifact of truncating the perturbation series 
and thereby neglecting higher-order nonlinear terms. 
Again, further simulations could help to resolve the issue.  

Other workers have investigated many-body interactions in charged colloids.  
Schmitz~\cite{Schmitz99} has developed a theory that describes sharing of 
counterions between macroions, analogous to molecular chemical bonding, 
and used the theory to study the influence of many-body effects on 
counterion distributions and the structure of colloidal crystals.
Sear~\cite{Sear00}, exploring a phenomenological model, showed that 
interactions among particles with internal degrees of freedom are, 
in general, non-pairwise-additive, and that triplet interactions may be 
attractive at the same time that pair interactions are repulsive.
Recently, Russ {\it et al.}~\cite{Russ02} 
solved the nonlinear PB equation for triplets of macroions 
immersed in an electrolyte.  Their conclusion that three-body effects are 
always cohesive agrees qualitatively with our results, and those of 
ref.~\cite{LA}.  We note, however, that three-body contributions 
to the grand potential, calculated in ref.~\cite{Russ02}, are not 
directly comparable to three-body interactions in the effective Hamiltonian, 
calculated here and in the simulations of Tehver {\it et al.}~\cite{Tehver}.  
In another study, Wu {\it et al.}~\cite{Wu00} extracted three-body forces 
from Monte Carlo simulations of macroion triplets in equilateral configurations.
These authors found attractive electrostatic three-body forces, but also 
detected a significant repulsive contribution attributable to hard-sphere
collisions between macroions and microions, which were modelled in the
simulations as charged hard spheres.  
Future extension of the response theory from point microions to hard-core
microions would allow a more direct comparison with these Monte Carlo data.

Influences on thermodynamic phase behavior of nonlinear corrections to
both effective interactions and the volume energy are now being explored.
It may be anticipated that these corrections will be especially significant 
for deionized suspensions of highly charged macroions.  Preliminary 
calculations of free energies and phase diagrams~\cite{Denton3} indicate 
that the spinodal-instability mechanism proposed to describe phase 
separation~\cite{vRH,vRDH} remains qualitatively valid -- at least under 
the assumption of fixed macroion charge -- but that nonlinearity can 
significantly shift the phase boundaries and, in some cases, 
{\it enhance} the tendency toward phase separation.

The role of nonlinear response and of effective many-body interactions 
in dense electron-ion (metallic) systems has long been recognized 
and discussed~\cite{Pethick70,Brovman70,Singh73,Rasolt75,Louis98}. 
In this context, it has been argued that nonlinear corrections to pair 
potentials and structure factors either are weak~\cite{Louis98} or can be 
incorporated into the linear response scheme~\cite{Rasolt75}, but that 
nonlinear corrections to the volume energy and thermodynamic properties
({\it e.g.}, bulk modulus) are more significant~\cite{Singh73}.  
Whether the same argument applies also to colloidal systems is being 
investigated in ongoing studies of phase behavior~\cite{Denton3}. 


\section{Summary and Conclusions}\label{Conclusions}
In summary, by incorporating nonlinear microion screening into a mean-field
response theory of charged colloids in the primitive model, we have
derived nonlinear corrections to the effective electrostatic interactions
between hard spherical macroions in bulk colloidal suspensions.
The key physical concept is that nonlinear screening entails both
effective many-body interactions {\it and} essential corrections to 
the effective pair potential and the one-body volume energy.  Effective 
triplet interactions are predicted to be always attractive, consistent 
with previous work~\cite{LA,Russ02}.  The effective pair potential
$v_{\rm eff}^{(2)}(r)$, which in the linear (DLVO) theory has screened-Coulomb 
form, is shortened in range by the influence of nonlinear screening, 
but remains purely repulsive within the physically reasonable range of 
renormalized macroion charges.  Predictions for $v_{\rm eff}^{(2)}(r)$ 
are in essentially perfect agreement with available {\it ab initio} 
simulation data~\cite{Tehver}.  The theory also predicts that triplet forces
are negligible between weakly charged macroions, consistent with
simulation~\cite{Tehver}, but can be significant for higher macroion charges.
Further simulations of more highly charged and concentrated macroions 
are now required to more severely test the theory.
 
Analytical and numerical results confirm that nonlinear effects become
qualitatively stronger with increasing macroion charge, increasing
macroion concentration, and decreasing salt concentration.  
In the dilute limit of zero macroion concentration, but nonzero salt
concentration, leading-order nonlinear corrections vanish.  Perhaps
the most practical application of the response theory, illustrated here,
is to mapping out the parameter ranges 
within which linear theories can be trusted.
Future work will explore implications of nonlinear screening for
thermodynamic properties, in particular, the phenomenon of phase separation 
at low salt concentrations and the stability of deionized charged
colloidal crystals.

\begin{acknowledgments} 
Helpful discussions and correspondence with Juan Antonio Anta, Neil Ashcroft, 
Jayanth Banavar, Carl Franck, Christos Likos, Hartmut L\"owen, 
Kenneth Schmitz, and Hao Wang are gratefully acknowledged.  
This work was supported by the National Science Foundation under Grant 
Nos.~DMR-0204020 and EPS-0132289.
\end{acknowledgments} 


\appendix*
\section{Comparison with Related Theoretical Approaches}
\subsection{Response Theory vs. Poisson-Boltzmann Theory}
\label{PB}
We demonstrate here that response theory, when combined with the 
random phase approximation (RPA) for the microion response functions 
(see Sec.~\ref{RPA}) is formally equivalent to Poisson-Boltzmann theory.
The ensemble-averaged number density profile of positive microions, in 
the presence of the macroion potential $\phi_{\rm ext}({\bf r})$, is given 
exactly by the Euler-Lagrange equation~\cite{HM,note5}, 
\begin{equation}
\rho_+({\bf r})=\frac{e^{\beta\mu_+}}{\Lambda_+^3}\exp\left[-\beta
ze\phi_{\rm ext}({\bf r})+c_+^{(1)}({\bf r};[\rho_+({\bf r}),\rho_-({\bf r})])
\right],
\label{appa-rho+r1}
\end{equation}
which follows from minimization of the grand potential functional with respect 
to $\rho_+({\bf r})$.  In Eq.~(\ref{appa-rho+r1}), $\mu_+$ and $\Lambda_+$
denote the chemical potential and thermal de Broglie wavelength of the 
positive microions, $\phi_{\rm ext}({\bf r})$ 
is the ``external" electrostatic potential of the macroions, and 
$c_+^{(1)}({\bf r};[\rho_+({\bf r}),\rho_-({\bf r})])$ is the one-particle 
direct correlation function (DCF) of the positive microions, which is
a functional of the inhomogeneous microion densities.  
Expanding $c_+^{(1)}({\bf r};[\rho_+({\bf r}),\rho_-({\bf r})])$ in a 
functional Taylor series about the average (bulk) microion densities, 
$n_+$ and $n_-$, we have 
\begin{eqnarray}
\rho_+({\bf r})&=&\frac{e^{\beta\mu_+}}{\Lambda_+^3}\exp\left[-\beta ze
\phi_{\rm ext}({\bf r})+c_+^{(1)}(n_+,n_-)+\int{\rm d}{\bf r}'\,
c_{++}^{(2)}(|{\bf r}-{\bf r}'|;n_+,n_-) \left[\rho_+({\bf r'})-n_+\right] 
\right.
\nonumber \\
&+&\left.\int{\rm d}{\bf r}'\, c_{+-}^{(2)}(|{\bf r}-{\bf r}'|;n_+,n_-)
\left[\rho_-({\bf r'})-n_-\right] + \cdots \right],
\label{appa-rho+r2}
\end{eqnarray}
where $c_{ij}^{(2)}(r;n_+,n_-)$, $i,j=\pm$, are the bulk microion two-particle 
DCFs, which are related to the one-particle DCFs via
\begin{equation}
c_{ij}^{(2)}(|{\bf r}-{\bf r}'|;n_+,n_-)=\lim_{\rho_{\pm}({\bf r})\to n_{\pm}}
\left[\frac{\delta c_i^{(1)}({\bf r};[\rho_+({\bf r}),\rho_-({\bf r})])}
{\delta\rho_j({\bf r}')}\right].
\label{c2}
\end{equation}

We now make the mean-field random phase approximation~\cite{HM}:
(1) neglecting three-particle and higher-order correlations,
{\it i.e.}, truncating the series in Eq.~(\ref{appa-rho+r2}),
and (2) ignoring short-range pair correlations by simply equating 
$c_{ij}^{(2)}(r;n_+,n_-)$ to their asymptotic long-range limits,
\begin{equation}
c_{ij}^{(2)}(r;n_+,n_-)\simeq -\beta v_{ij}(r), \qquad i,j=\pm.
\label{appa-rpa}
\end{equation}
Equation~(\ref{appa-rho+r2}) then becomes
\begin{eqnarray}
\rho_+({\bf r})&\simeq&n_+\exp\left[-\beta\left(
ze\phi_{\rm ext}({\bf r})+\int{\rm d}{\bf r}'\,
v_{++}(|{\bf r}-{\bf r}'|) \left[\rho_+({\bf r'})-n_+\right] 
\right.\right.
\nonumber \\
&&\hspace{1cm}\left.\left.+\int{\rm d}{\bf r}'\,
v_{+-}(|{\bf r}-{\bf r}'|) \left[\rho_-({\bf r'})-n_-\right] 
\right)\right]
\nonumber \\
&=&n_+\exp[-\beta ze\phi({\bf r})],
\label{appa-rho+r}
\end{eqnarray}
where we have used $n_+=(e^{\beta\mu_+}/\Lambda_+^3)\exp[c_+^{(1)}(n_+,n_-)]$
and have identified
\begin{equation}
\phi({\bf r})=\phi_{\rm ext}({\bf r})+\int{\rm d}{\bf r}'\,
\frac{ze}{\epsilon|{\bf r}-{\bf r}'|}
\left[\rho_+({\bf r}')-n_+-\rho_-({\bf r}')+n_-\right]
\label{appa-phi2}
\end{equation}
as the {\it total} electrostatic potential, due to both the macroions 
and the surrounding microions.
Similarly, the density profile of negative microions is given by
\begin{eqnarray}
\rho_-({\bf r})&\simeq&n_-\exp\left[-\beta\left(
-ze\phi_{\rm ext}({\bf r})+\int{\rm d}{\bf r}'\,
v_{+-}(|{\bf r}-{\bf r}'|) \left[\rho_+({\bf r'})-n_+\right] 
\right.\right.
\nonumber \\
&&\left.\left.\hspace{1cm}+\int{\rm d}{\bf r}'\,
v_{--}(|{\bf r}-{\bf r}'|) \left[\rho_-({\bf r'})-n_-\right] 
\right)\right]
\nonumber \\
&=&n_-\exp[\beta ze\phi({\bf r})].
\label{appa-rho-r}
\end{eqnarray}
Combining Eqs.~(\ref{appa-rho+r}) and (\ref{appa-rho-r}) with the 
Poisson equation 
\begin{equation}
\nabla^2\phi({\bf r})=-\frac{4\pi}{\epsilon}\left\{ze\rho_+({\bf r})
-ze\rho_-({\bf r})\right\},
\label{Poisson}
\end{equation}
we obtain
\begin{equation}
\nabla^2\phi({\bf r})
=-\frac{4\pi ze}{\epsilon}\left\{n_+\exp[-\beta ze\phi({\bf r})]
-n_-\exp[\beta ze\phi({\bf r})]\right\},
\label{PB-salt}
\end{equation}
which is the PB equation for macroions in a symmetric $z:z$ electrolyte.
We conclude that the RPA-based response theory is formally equivalent 
to the mean-field PB theory.  This is not surprising, given that both
approaches neglect microion correlations.  Response theory, however, 
provides a powerful framework for going beyond a mean-field description 
by systematically including microion correlations via more accurate 
approximations for the response functions of the microion plasma.

\subsection{Response Theory vs. Integral Equation Theory}
\label{HNC}
Here we show that linear response theory is equivalent to a linearized
hypernetted-chain (HNC) approximation in integral-equation theory.
Substituting Eqs.~(\ref{chi++r}-\ref{chi--r}) into the Fourier transform of 
Eq.~(\ref{rho+-k}), the linear response expressions for the ensemble-averaged 
number density profiles of positive and negative microions are 
\begin{equation}
\frac{\rho_+({\bf r})}{n_+}=1-\beta\sum_i\left[v_{m+}(|{\bf r}-{\bf R}_i|)
+\int{\rm d}{\bf r}'\,[n_+h_{++}(|{\bf r}-{\bf r}'|)
-n_-h_{+-}(|{\bf r}-{\bf r}'|)] v_{m+}(|{\bf r}'-{\bf R}_i|)
\right]
\label{appb-rho+r1}
\end{equation}
and
\begin{equation}
\frac{\rho_-({\bf r})}{n_-}=1+\beta\sum_i\left[v_{m+}(|{\bf r}-{\bf R}_i|)
+\int{\rm d}{\bf r}'\,[n_-h_{--}(|{\bf r}-{\bf r}'|)
-n_+h_{+-}(|{\bf r}-{\bf r}'|)] v_{m+}(|{\bf r}'-{\bf R}_i|)
\right].
\label{appb-rho-r1}
\end{equation}
On the other hand, from Eq.~(\ref{appa-rho+r2}), we have the exact relations
\begin{eqnarray}
\rho_+({\bf r})&=&n_+\exp\left[-\beta\sum_i v_{m+}(|{\bf r}-{\bf R}_i|)
+\int{\rm d}{\bf r}'\, c_{++}^{(2)}(|{\bf r}-{\bf r}'|)
\left[\rho_+({\bf r'})-n_+\right] \right.
\nonumber \\
&+&\left.\int{\rm d}{\bf r}'\, c_{+-}^{(2)}(|{\bf r}-{\bf r}'|)
\left[\rho_-({\bf r'})-n_-\right] + \cdots \right]
\label{appb-rho+r2}
\end{eqnarray}
and
\begin{eqnarray}
\rho_-({\bf r})&=&n_-\exp\left[\beta\sum_i v_{m+}(|{\bf r}-{\bf R}_i|)
+\int{\rm d}{\bf r}'\, c_{--}^{(2)}(|{\bf r}-{\bf r}'|)
\left[\rho_-({\bf r'})-n_-\right] \right.
\nonumber \\
&+&\left.\int{\rm d}{\bf r}'\, c_{+-}^{(2)}(|{\bf r}-{\bf r}'|)
\left[\rho_+({\bf r'})-n_+\right] + \cdots \right].
\label{appb-rho-r2}
\end{eqnarray}
Truncating the expansions on the right side of Eqs.~(\ref{appb-rho+r2}) 
and (\ref{appb-rho-r2}) at the level of two-particle correlations amounts 
to the HNC approximation in integral-equation theory.  
If we further linearize the exponential functions (expanding and neglecting 
all but the first two terms), substitute recursively for $\rho_+({\bf r})$ 
and $\rho_-({\bf r})$, and use the Ornstein-Zernike (OZ) relation for 
mixtures~\cite{HM}
\begin{equation}
h_{ij}(r)=c^{(2)}_{ij}(r)+\sum_k n_k\int{\rm d}{\bf r}'\, 
c^{(2)}_{ik}(|{\bf r}-{\bf r}'|)h_{kj}(r'),
\end{equation}
we recover Eqs.~(\ref{appb-rho+r1}) and (\ref{appb-rho-r1}).
Thus, the linear response approximation is equivalent to a linearized-HNC 
closure of the OZ relation, while nonlinear response generates 
new closures.


\newpage




\unitlength1mm

\begin{figure}
\includegraphics[width=0.55\columnwidth]{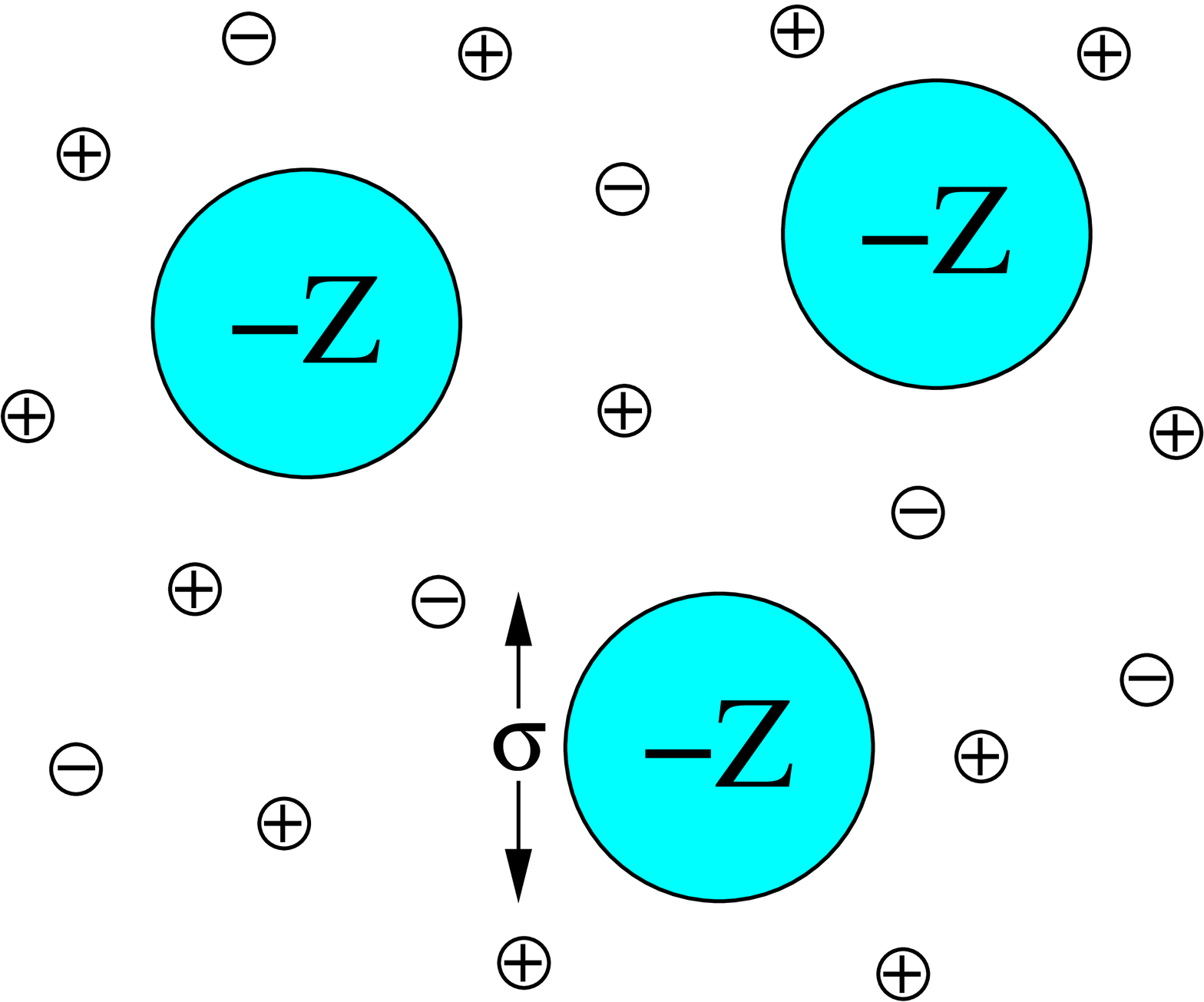}
\caption{\label{fig-model} Primitive model of a charged colloidal suspension: 
hard macroions (valence $-Z$ and diameter $\sigma$) and point microions 
(counterions and salt ions) in a dielectric continuum (not shown).}
\end{figure}

\begin{figure}
\includegraphics[width=0.55\columnwidth]{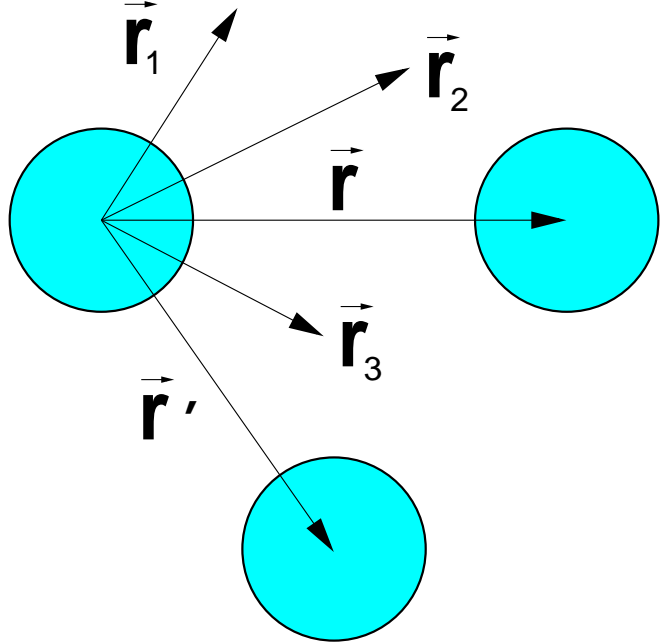}
\caption{\label{fig-diagram} Geometry for the physical interpretation
of response theory (see Sec.~\ref{interpretation}).  Vectors ${\bf r}$ 
and ${\bf r}'$ define center-to-center displacements of macroions.
Vectors ${\bf r}_1$, ${\bf r}_2$, and ${\bf r}_3$ define points
at which either the macroion ``external" potential acts or a change
in microion density is induced.
}
\end{figure}

\begin{figure}
\includegraphics[width=0.55\columnwidth]{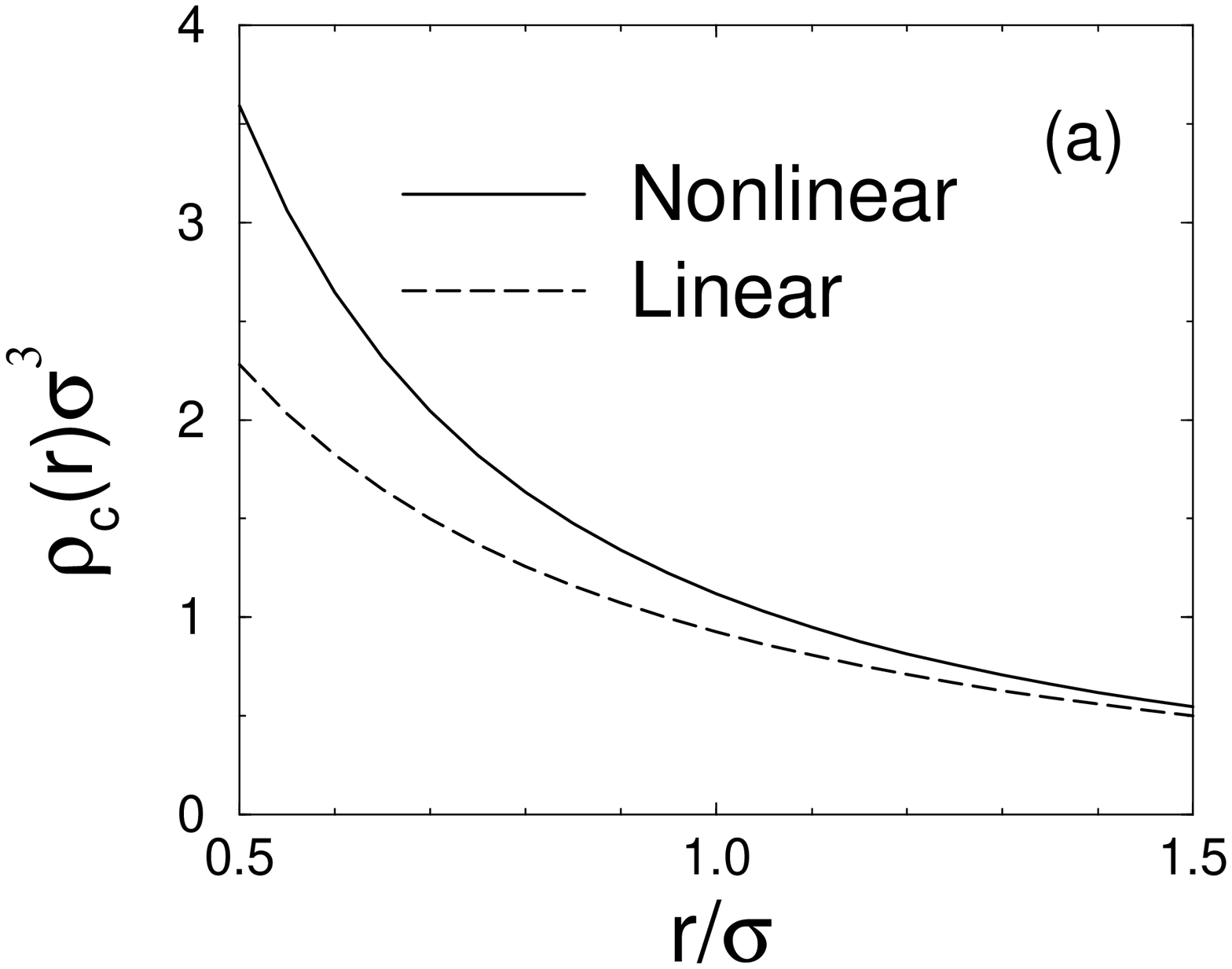} \\[10ex]
\includegraphics[width=0.55\columnwidth]{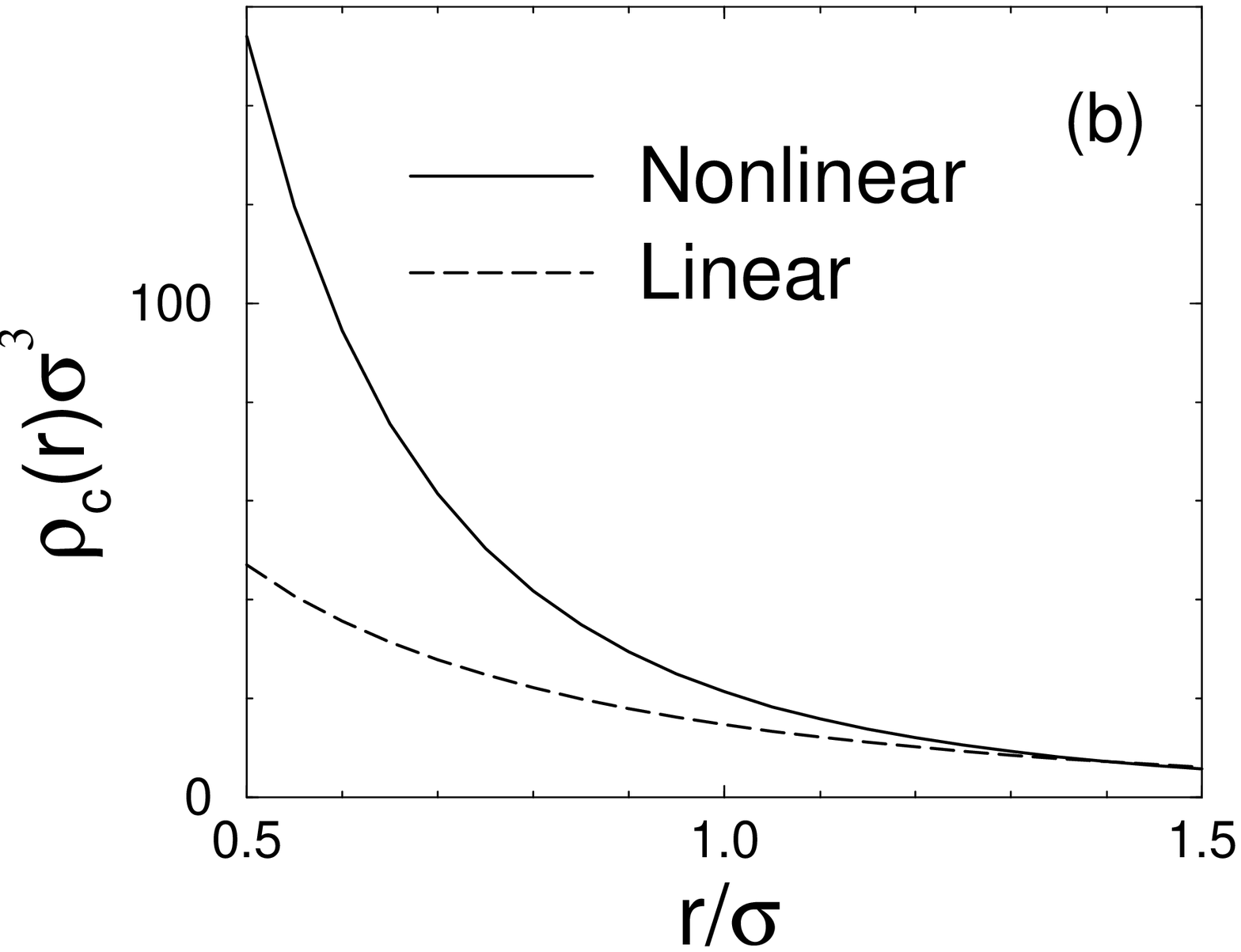}
\caption{\label{fig-rhocr} Ensemble-averaged counterion density around a 
single macroion for macroion diameter $\sigma=100$ nm and valence (a) $Z=100$,
(b) $Z=500$, at volume fraction $\eta=0.01$ and zero salt concentration. 
Dashed curves: linear response theory.
Solid curves: nonlinear response theory (first-order correction).} 
\end{figure}

\begin{figure}
\includegraphics[width=0.55\columnwidth]{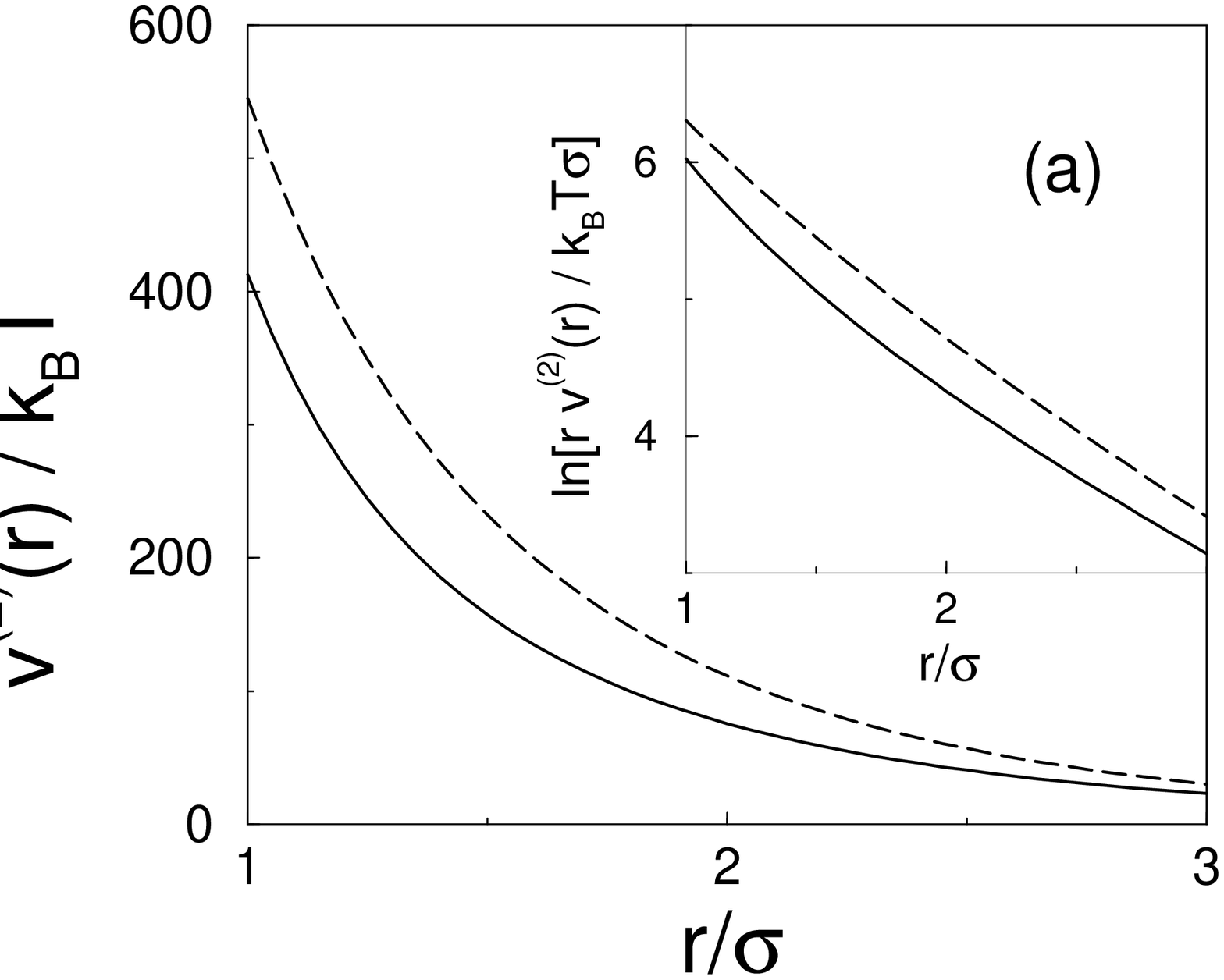} \\[10ex]
\includegraphics[width=0.55\columnwidth]{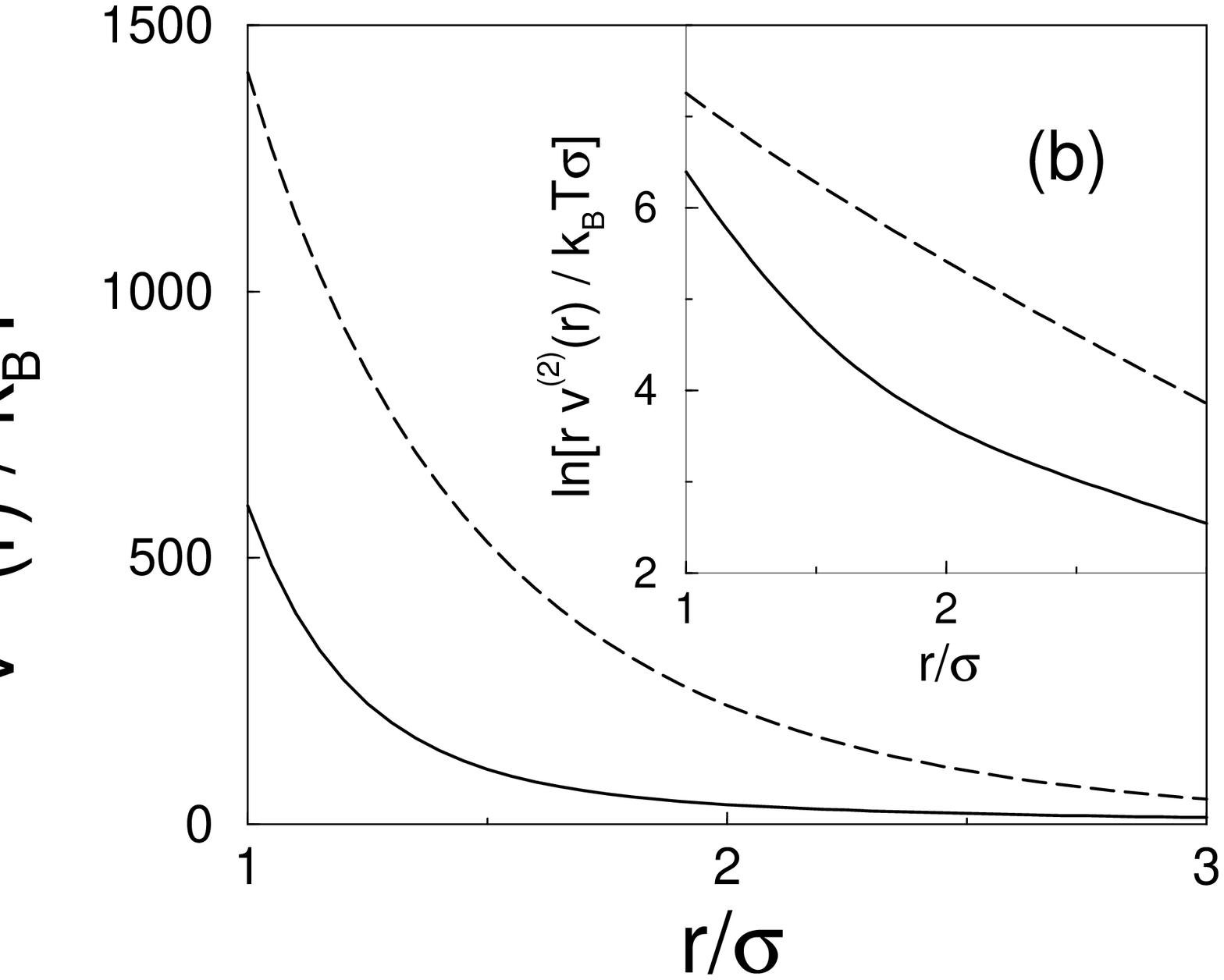}
\end{figure}

\begin{figure}
\includegraphics[width=0.55\columnwidth]{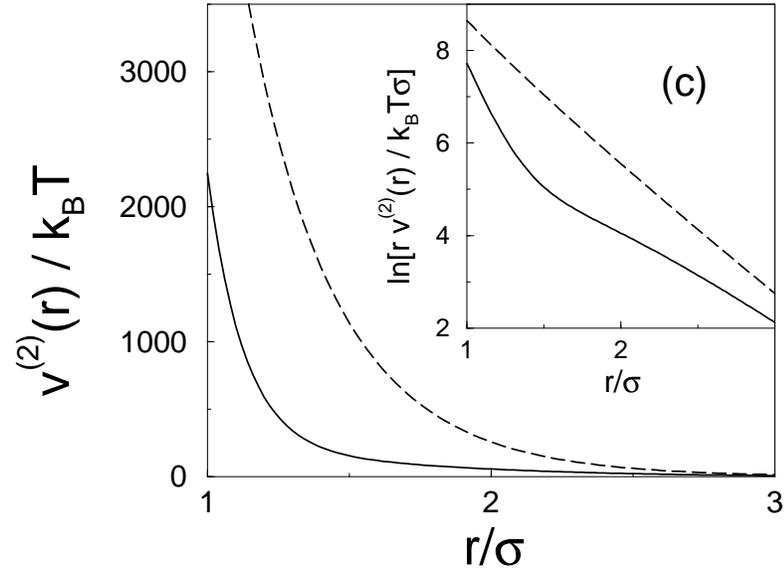}
\caption{\label{fig-v2r} Effective pair interactions for 
macroion diameter $\sigma$, valence $Z$, volume fraction $\eta$, 
and salt concentration $c_s$:
(a) $\sigma=100$ nm, $Z=400$, $\eta=0.01$, $c_s=1$ $\mu$M;
(b) $\sigma=100$ nm, $Z=700$, $\eta=0.01$, $c_s=1$ $\mu$M;
(c) $\sigma=652$ nm, $Z=5000$, $\eta=0.0352$, $c_s=0.2$ $\mu$M
(chosen to compare with ref.~\cite{Grier2}).
Dashed curves: linear response prediction.
Solid curves: nonlinear response prediction. 
The insets show that for sufficiently weakly charged macroions
(a) the effective pair interaction may be fit by a Yukawa potential
(with a lower effective charge), while for more highly charged macroions 
(b,c) deviations from linear behavior can be significant.}
\end{figure}

\begin{figure}
\includegraphics[width=0.55\columnwidth]{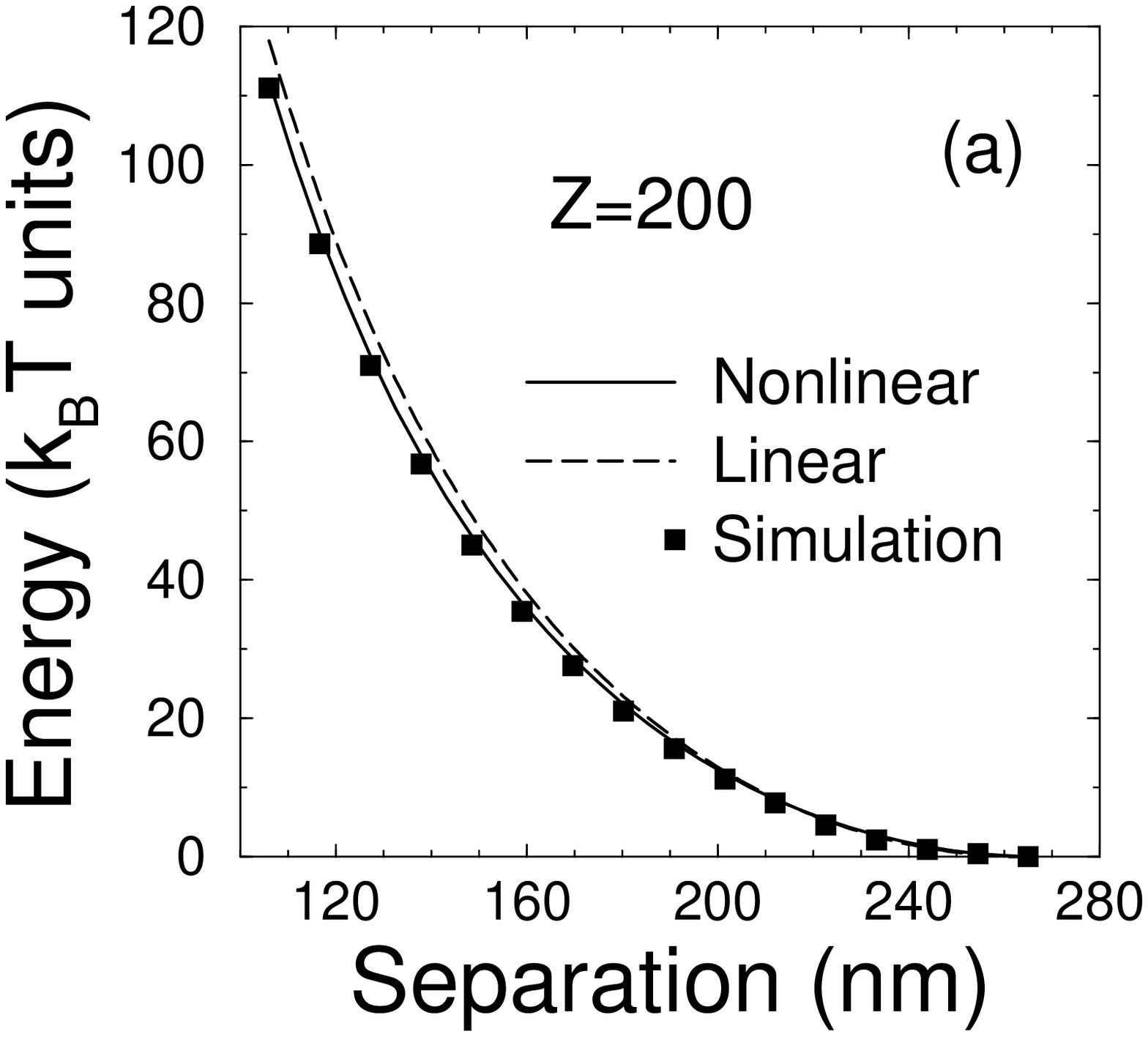} \\[5ex]
\includegraphics[width=0.55\columnwidth]{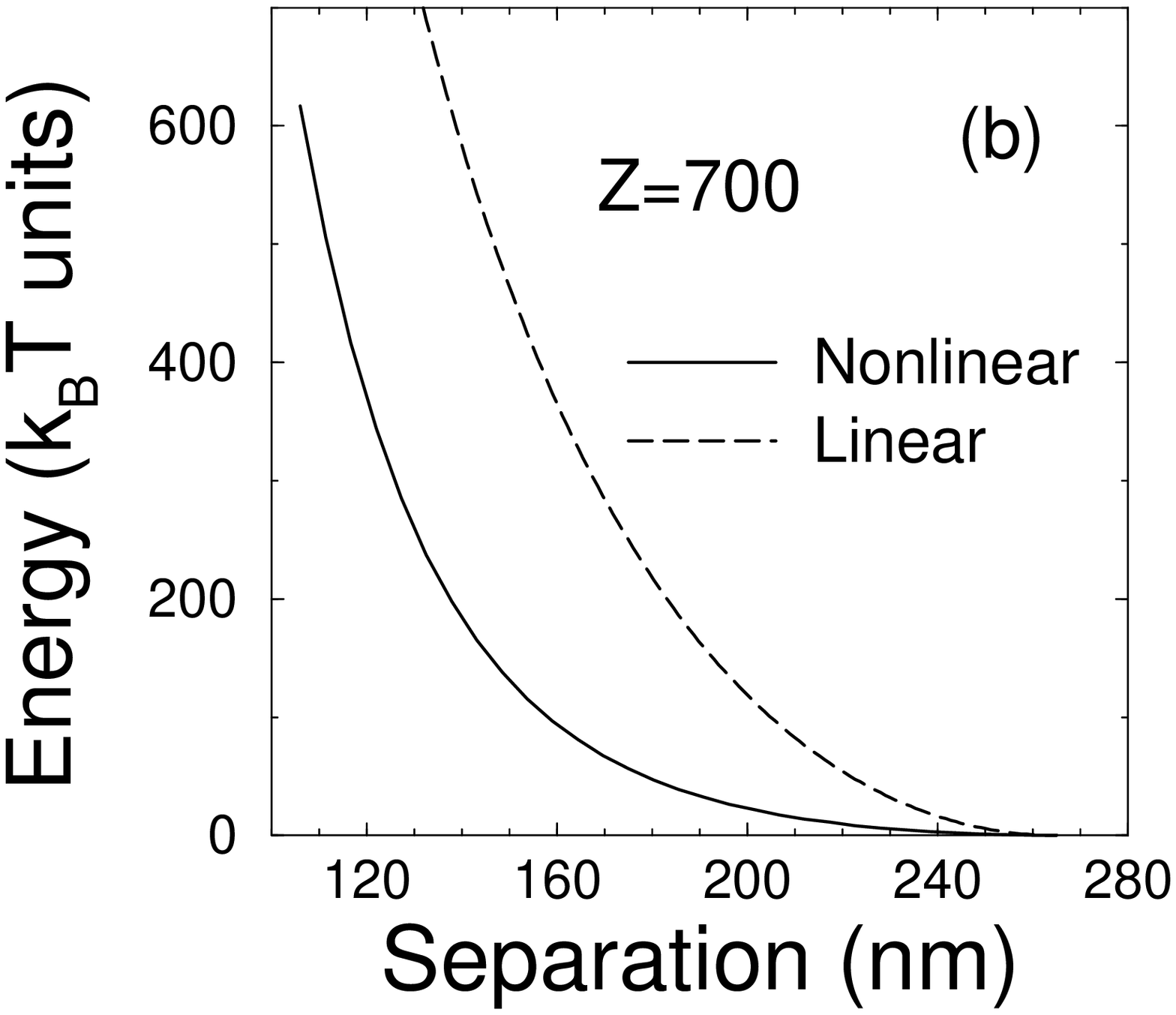}
\caption{\label{fig-epot} Total interaction potential energy for two macroions,
of diameter $\sigma=106$ nm and valence (a) $Z=200$, (b) $Z=700$,
in a cubic box of length 530 nm with periodic boundary conditions
at zero salt concentration.
The potentials are shifted to zero at maximum macroion separation.
Dashed curves: linear response prediction. 
Solid curves: nonlinear response prediction.
Symbols: {\it ab initio} simulation data~\cite{Tehver}.} 
\end{figure}

\begin{figure}
\includegraphics[width=0.55\columnwidth]{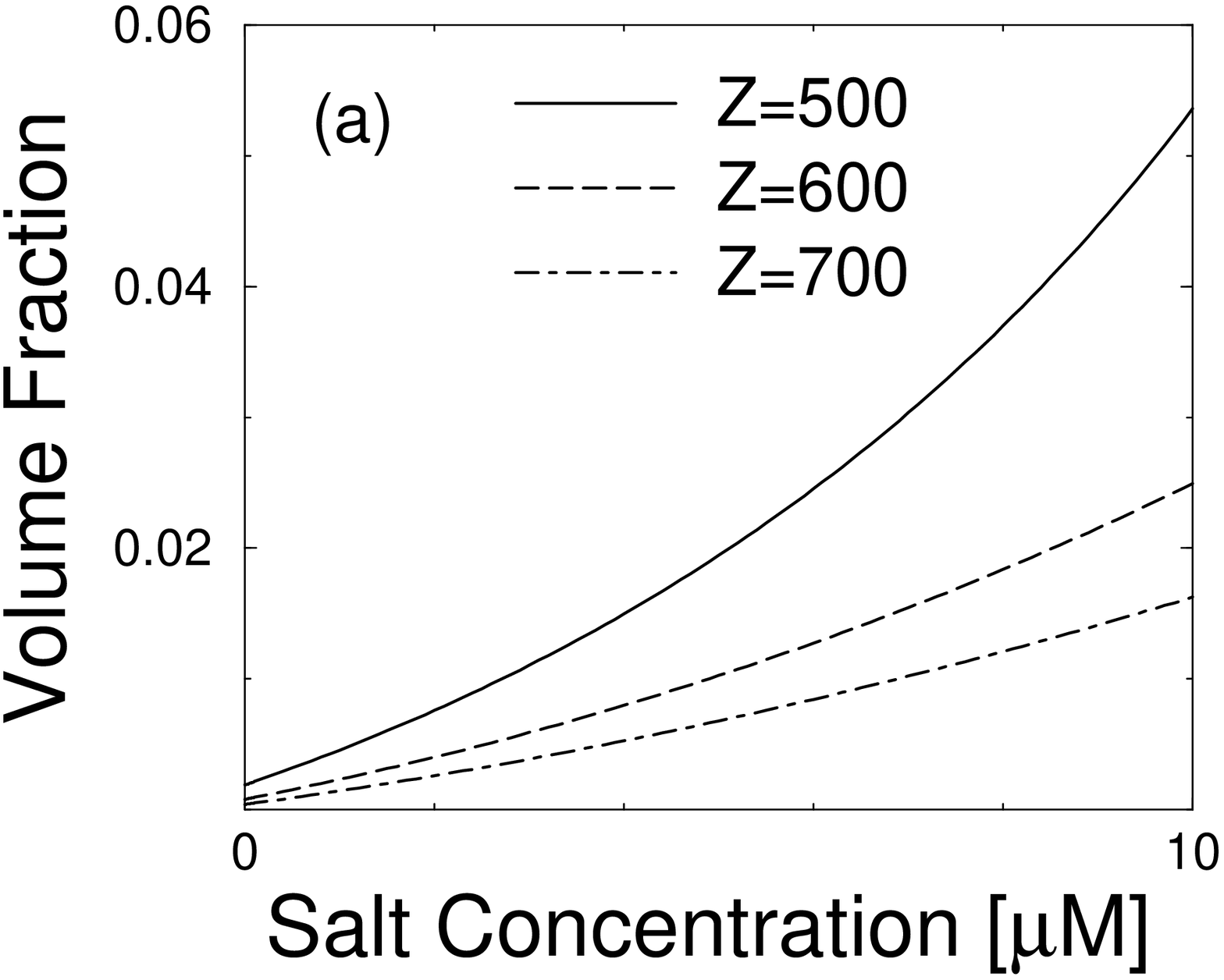} \\[5ex]
\includegraphics[width=0.55\columnwidth]{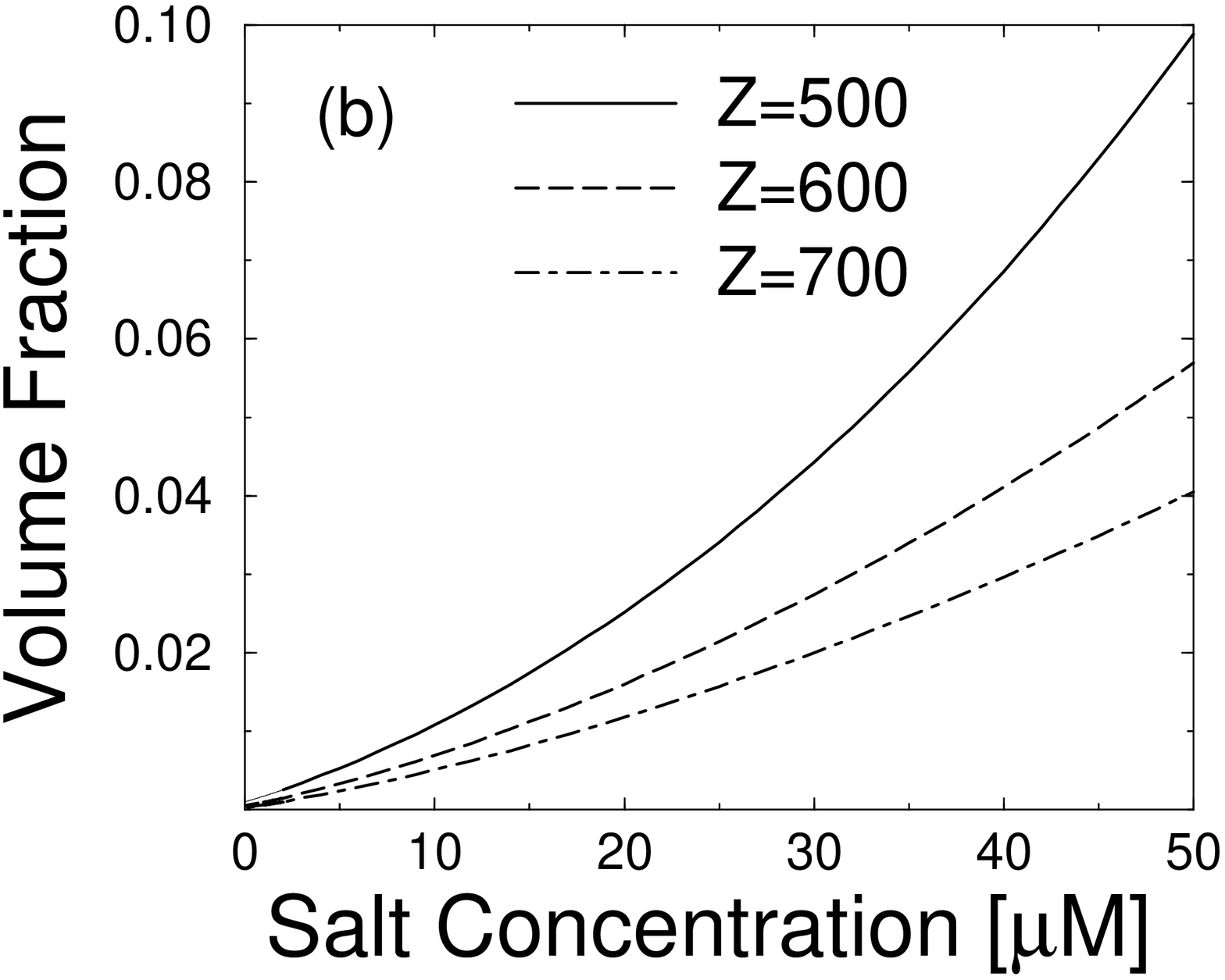}
\caption{\label{fig-vscan} Map of nonlinear deviations from linear response
theory for 
macroions of diameter $\sigma=100$ nm and valences, from top to bottom, 
$Z=500, 600, 700$, for fcc crystal structure.  Systems with macroion volume 
fractions $\eta$ and salt concentrations $c_s$ above the respective curves 
deviate from the linear response pair potential at the fcc nearest-neighbor 
distance by at least (a) $1~k_BT$ or (b) $0.1~k_BT$.}
\end{figure}

\begin{figure}
\includegraphics[width=0.55\columnwidth]{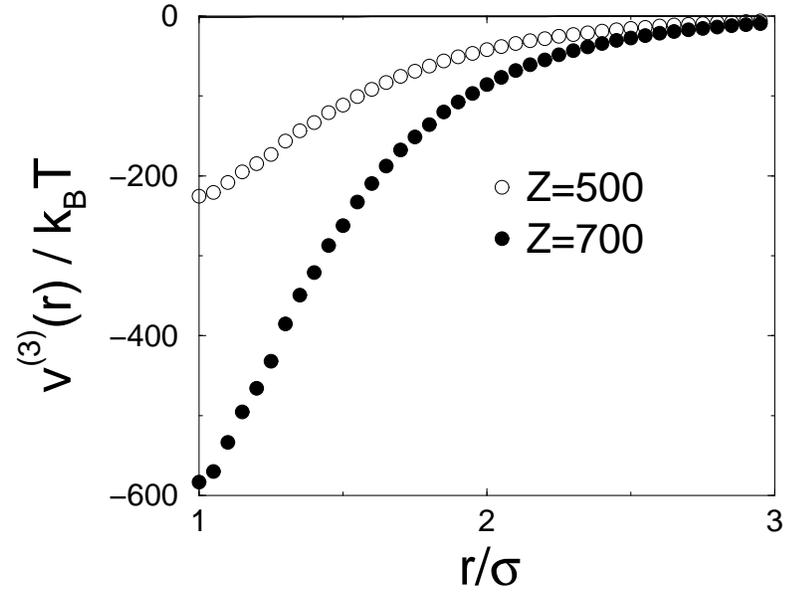}
\caption{\label{fig-v3r} Effective three-body interaction between 
three macroions, arranged in an equilateral triangle of side length $r$, 
with macroion diameter $\sigma=100$ nm, 
valence $Z=500$ (open circles), $Z=700$ (filled circles),
volume fraction $\eta=0.01$, and salt concentration $c_s=1$ $\mu$M.
Computed by Monte Carlo integration of Eq.~(\ref{v3effr-salt}),
with numerical errors comparable to symbol size.}
\end{figure}

\begin{figure}
\includegraphics[width=0.55\columnwidth]{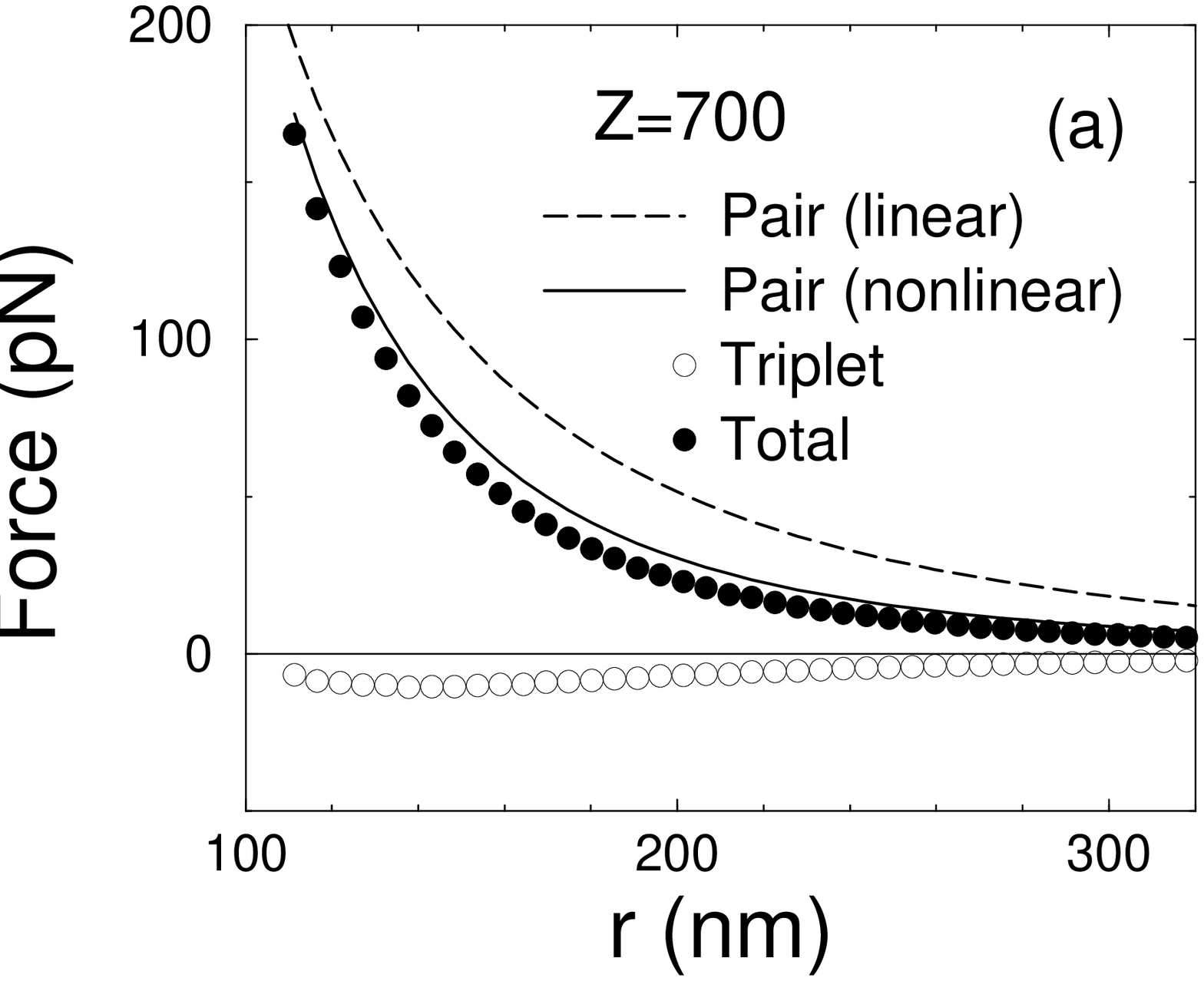} \\[10ex]
\includegraphics[width=0.55\columnwidth]{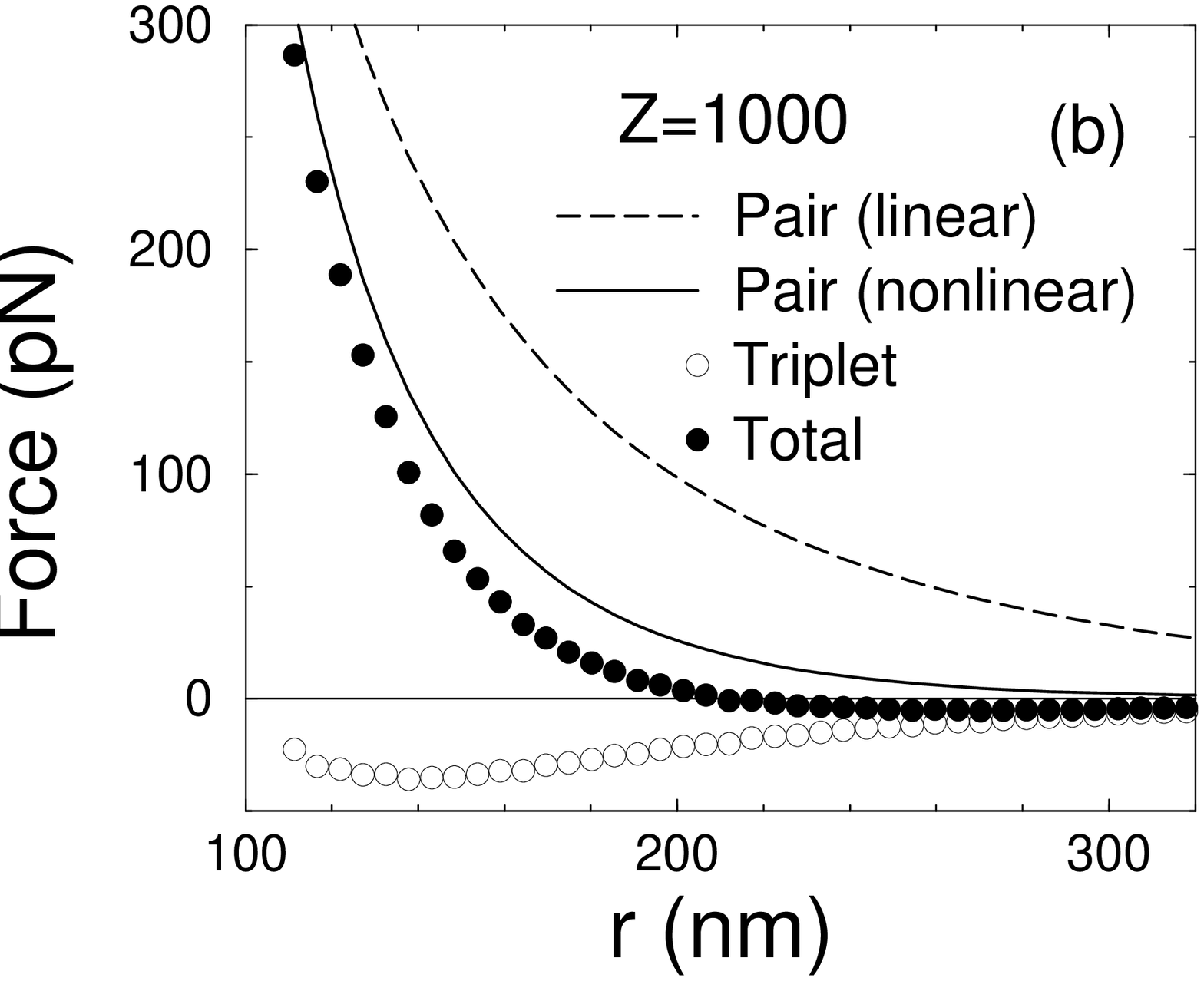}
\caption{\label{fig-f3r} Predicted force on a macroion in an 
equilateral-triangle arrangement of three macroions, 
each of diameter $\sigma=106$ nm and 
valence (a) $Z=700$, (b) $Z=1000$, in a cubic box of length 1000 nm 
with periodic boundary conditions at zero salt concentration.
Dashed curves: sum of linear response effective pair forces.
Solid curves: sum of nonlinear effective pair forces.
The symbols are theoretical predictions
for the effective triplet force (open circles) and the total 
(pair plus triplet) effective force (filled circles).} 
\end{figure}


\newpage
\begin{references}


\bibitem{Hunter}
R.~J.~Hunter, {\it Foundations of Colloid Science}
(Oxford University Press, Oxford, 1986).
 
\bibitem{Pusey}
P.~N.~Pusey, in {\it Liquids, Freezing and Glass Transition},
session 51, ed. J.-P.~Hansen, D.~Levesque, and J.~Zinn-Justin
(North-Holland, Amsterdam, 1991).

\bibitem{Schmitz}
K.~S.~Schmitz, {\it Macroions in Solution and Colloidal Suspension}
(VCH, New York, 1993).

\bibitem{PE1}
F.~Oosawa, {\it Polyelectrolytes} (Dekker, New York, 1971).

\bibitem{PE2}
{\it Polyelectrolytes}, ed. M.~Hara (Dekker, New York, 1993).

\bibitem{Tata}
B.~V.~R.~Tata, M.~Rajalakshmi, and A.~K.~Arora, \PRL {\bf 69}, 3778 (1992).

\bibitem{Ise}
K.~Ito, H.~Yoshida, and N.~Ise, {\it Science} {\bf 263}, 66 (1994).

\bibitem{Matsuoka}
H.~Matsuoka, T.~Harada, and H.~Yamaoka, {\it Langmuir} {\bf 10}, 4423(1994); 
H.~Matsuoka, T.~Harada, K.~Kago, and H.~Yamaoka, {\it ibid} {\bf 12}, 
5588 (1996); T.~Harada, H.~Matsuoka, T.~Ikeda, and H.~Yamaoka, {\it ibid} 
{\bf 15}, 573 (1999).

\bibitem{Groehn00}
F.~Gr\"ohn and M.~Antonietti, {\it Macromolecules} {\bf 33}, 5938 (2000).

\bibitem{Grier2}
A.~E.~Larsen and D.~G.~Grier, {\it Nature} {\bf 385}, 230 (1997).

\bibitem{Fraden}
G.~M.~Kepler and S.~Fraden, \PRL {\bf 73}, 356 (1994).

\bibitem{Grier1}
A.~E.~Larsen and D.~G.~Grier, \PRL {\bf 76}, 3862 (1996);
J.~C.~Crocker and D.~G.~Grier, \PRL {\bf 77}, 1897 (1996).

\bibitem{Levin02}
Y.~Levin, {\it Rep.~Prog.~Phys.} {\bf 65}, 1577 (2002).

\bibitem{Likos01}
C.~N.~Likos, {\it Phys.~Rep.} {\bf 348}, 267 (2001).

\bibitem{Belloni00}
L.~Belloni, \JPCM {\bf 12}, R549 (2000).

\bibitem{Hansen-Lowen}
J.-P.~Hansen and H.~L\"owen, {\it Ann.~Rev.~Phys.~Chem.} {\bf 51}, 209 (2000).

\bibitem{Robbins88}
M.~O.~Robbins, K.~Kremer, and G.~G.~Grest, \JCP {\bf 88}, 3286 (1988).

\bibitem{Meijer91}
E.~J.~Meijer and D.~Frenkel, \JCP {\bf 94}, 2269 (1991).

\bibitem{Auer02}
S.~Auer and D.~Frenkel, \JPCM {\bf 14}, 7667 (2002).

\bibitem{Hynninen03} 
A.-P.~Hynninen and M.~Dijkstra, \PR E {\bf 68}, 21407 (2003).

\bibitem{Tehver}
R.~Tehver, F.~Ancilotto, F.~Toigo, J.~Koplik, and J.~R.~Banavar,
\PR E {\bf 59}, R1335 (1999).

\bibitem{Lowen}
H.~L\"owen, P.~A.~Madden, and J.-P.~Hansen, \PRL {\bf 68}, 1081 (1992);
H.~L\"owen, J.-P.~Hansen, and P.~A.~Madden, \JCP {\bf 98}, 3275 (1993);
H.~L\"owen and G.~Kramposthuber, \EPL {\bf 23}, 673 (1993).

\bibitem{Stevens96}
M.~J.~Stevens, M.~L.~Falk, and M.~O.~Robbins, \JCP {\bf 104}, 5209 (1996).

\bibitem{Linse-Lobaskin}
P.~Linse, \JCP {\bf 110}, 3493 (1999);
V.~Lobaskin and P.~Linse, \JCP {\bf 111}, 4300 (1999);
P.~Linse and V.~Lobaskin, \PRL {\bf 83}, 4208 (1999);
P.~Linse, \JCP {\bf 113}, 4359 (2000);
J.~Re\v{s}\v{c}i\v{c} and P.~Linse, \JCP {\bf 114}, 10131 (2001);
V.~Lobaskin, A.~Lyubartsev, and P.~Linse, \PR E {\bf 63}, 20401 (2001);
V.~Lobaskin and K.~Qamhieh, {\it J.~Phys.~Chem.} B {\bf 107}, 8022 (2003).

\bibitem{Holm}
R.~Messina, C.~Holm, and K.~Kremer, \PRL {\bf 85}, 872 (2000);
\EPJ E {\bf 4}, 363 (2001).

\bibitem{Damico}
E.~Allahyarov, I.~D'Amico, and H.~L\"owen, \PRL {\bf 81}, 1334 (1998).

\bibitem{DL}
B.~V.~Derjaguin and L.~Landau, {\it Acta Physicochimica} (USSR) {\bf 14},
633 (1941).

\bibitem{VO}
E.~J.~W.~Verwey and J.~T.~G.~Overbeek, {\it Theory of the Stability of
Lyophobic Colloids} (Elsevier, Amsterdam, 1948).

\bibitem{Graf}
H.~Graf and H.~L\"owen, \PR E {\bf 57}, 5744 (1998).

\bibitem{vRH}
R.~van Roij and J.-P.~Hansen, \PRL {\bf 79}, 3082 (1997).

\bibitem{vRDH}
R.~van Roij, M.~Dijkstra, and J.-P.~Hansen, \PR E {\bf 59}, 2010 (1999).

\bibitem{Silbert}
M.~J.~Grimson and M.~Silbert, \MP {\bf 74}, 397 (1991).

\bibitem{Denton1}
A.~R.~Denton, \JPCM {\bf 11}, 10061 (1999).

\bibitem{Denton2}
A.~R.~Denton, \PR E {\bf 62}, 3855 (2000).

\bibitem{Warren}
P.~B.~Warren, \JCP {\bf 112}, 4683 (2000).
 
\bibitem{Chan01}
D.~Y.~C.~Chan, P.~Linse, and S.~N.~Petris, {\it Langmuir} {\bf 17}, 4202 (2001).

\bibitem{Rouzina96}
I.~Rouzina and V.~A.~Bloomfield, \JPC {\bf 100}, 9977 (1996). 

\bibitem{Liu97}
B.-Y.~Ha and A.~J.~Liu, \PRL {\bf 79}, 1289 (1997); \PR E {\bf 58}, 6281 (1998).

\bibitem{Stevens99}
M.~J.~Stevens \PRL {\bf 82}, 101 (1999); {\it Biophys.~J.} {\bf 80}, 130 (2001).

\bibitem{Shklovskii99}
B.~I.~Shklovskii, \PR E {\bf 60}, 5802 (1999);
T.~T.~Nguyen, I.~Rouzina, and B.~I.~Shklovskii, \PR E {\bf 60}, 7032 (1999). 

\bibitem{Nguyen00}
T.~T.~Nguyen, I.~Rouzina, and B.~I.~Shklovskii, \JCP {\bf 112}, 2562 (2000);
T.~T.~Nguyen, A.~Yu.~Grosberg, and B.~I.~Shklovskii, \JCP {\bf 113}, 
1110 (2000).

\bibitem{Gelbart01}
W.~M.~Gelbart, R.~F.~Bruinsma, P.~A.~Pincus, and V.~A.~Parsegian, 
{\it Physics Today} {\bf 53} (Sept.~2000), p.~38.

\bibitem{Patey80}
G.~N.~Patey, \JCP {\bf 72}, 5763 (1980).

\bibitem{Belloni86}
L.~Belloni, \PRL {\bf 57}, 2026 (1986).

\bibitem{Khan87}
S.~Khan and D.~Ronis, \MP {\bf 60}, 637 (1987);
S.~Khan, T.~L.~Morton, and D.~Ronis, \PR A {\bf 35}, 4295 (1987).

\bibitem{Carbajal-Tinoco02}
M.~D.~Carbajal-Tinoco and P.~Gonz\'alez-Mozuelos, \JCP {\bf 117}, 2344 (2002).

\bibitem{Outhwaite02}
L.~B.~Bhuiyan and C.~W.~Outhwaite, \JCP {\bf 116}, 2650 (2002).

\bibitem{Petris02}
S.~N.~Petris and D.~Y.~C.~Chan, \JCP {\bf 116}, 8588 (2002).

\bibitem{Anta}
J.~A.~Anta and S.~Lago, \JCP {\bf 116}, 10514 (2002);
V.~Morales, J.~A.~Anta, and S.~Lago, {\it Langmuir} {\bf 19}, 475 (2003).

\bibitem{vonGrunberg01}
H.~H.~von Gr\"unberg, R.~van Roij, and G.~Klein \EPL {\bf 55}, 580 (2001).

\bibitem{Deserno02}
M.~Deserno and H.~H.~von Gr\"unberg, \PR E {\bf 66}, 011401 (2002).

\bibitem{Tamashiro03}
M.~N.~Tamashiro and H.~Schiessel, \JCP {\bf 119}, 1855 (2003).

\bibitem{Bowen-Sharif98}
W.~R.~Bowen and A.~O.~Sharif, {\it Nature} {\bf 393}, 663 (1998).

\bibitem{Gray99}
J.~J.~Gray, B.~Chiang, and R.~T.~Bonnecaze, {\it Nature} {\bf 402}, 750 (1999).

\bibitem{Dobnikar03}
J.~Dobnikar, R.~Rzehak, and H.~H.~von Gr\"unberg, \EPL {\bf 61}, 695 (2003);
J.~Dobnikar, Y.~Chen, R.~Rzehak, and H.~H.~von Gr\"unberg, \JPCM {\bf 15}, 
S263 (2003).

\bibitem{Goulding99}
D.~Goulding and J.-P.~Hansen, {\EPL} {\bf 46}, 407 (1999).

\bibitem{HGvR00}
J.-P.~Hansen, D.~Goulding, and R.~van Roij, {\it J.~Phys.~IV France} 
{\bf 10}, Pr5-27 (2000).

\bibitem{LA}
H.~L\"owen and E.~Allahyarov, \JPCM {\bf 10}, 4147 (1998).

\bibitem{Fisher94}
M.~E.~Fisher, \JSP {\bf 75}, 1 (1994); X.~Li, Y.~Levin, and M.~E.~Fisher,
\EPL {\bf 26}, 683 (1994);
M.~E.~Fisher, Y.~Levin, and X.~Li, \JCP {\bf 101}, 2273 (1994).

\bibitem{Phillies95}
N.~V.~Sushkin and G.~D.~J.~Phillies, \JCP {\bf 103}, 4600 (1995).

\bibitem{Gonzalez01}
L.~E.~Gonz\'alez, D.~J.~Gonz\'alez, M.~Silbert, and S.~Baer, \MP {\bf 99}, 
875 (2001).

\bibitem{Rowlinson84}
J.~S.~Rowlinson, \MP {\bf 52}, 567 (1984).

\bibitem{HM}
J.-P.~Hansen and I.~R.~McDonald, {\it Theory of Simple Liquids},
$2 ^{nd}$ ed.~(Academic, London, 1986).  

\bibitem{AS}
N.~W.~Ashcroft and D.~Stroud, {\it Solid State Phys.} {\bf 33}, 1 (1978).

\bibitem{Hafner87}
J.~Hafner, {\it From Hamiltonians to Phase Diagrams} (Springer, Berlin, 1987).

\bibitem{Ashcroft66}
N.~W.~Ashcroft, {\it Phys.~Lett.} {\bf 23}, 48 (1966).

\bibitem{note1}
We note that expansion about zero macroion charge ($u=0$) in 
Eq.~(\ref{delta-rho}) is here the only natural choice, thus 
avoiding ambiguities encountered in other linearization 
schemes~\cite{vonGrunberg01,Deserno02,Tamashiro03}.

\bibitem{Trigger}
E.~A.~Allahyarov, L.~I.~Podloubny, and P.~P.~J.~M.~Schram, and S.~A.~Trigger, 
{\it Physica} A {\bf 220}, 349 (1995).

\bibitem{note2}
We note that the counterion-counterion pair correlation function $h_{cc}(r)$ 
in Eq.~(\ref{hr}) also appears in the distribution function approach of 
Chan {\it et al}.~({\it cf}. Eq.~(45) of ref.~\cite{Chan01}).

\bibitem{note3}
Equation (\ref{E03}) has also been derived by a density-functional approach 
in refs.~\cite{Graf} and \cite{vRDH}.

\bibitem{GR}
I.~S.~Gradstein and I.~M.~Ryzhik, {\it Table of Integrals, Series, and
Products} (Academic, New York, 1980).

\bibitem{Klein02}
R.~Klein, H.~H.~von Gr\"unberg, C.~Bechinger, M.~Brunner, and
V.~Lobaskin, \JPCM {\bf 14}, 7631 (2002).

\bibitem{Franck}
R.~V.~Durand and C.~Franck, \PR E {\bf 61}, 6922 (2000).

\bibitem{Alexander}
S.~Alexander, P.~M.~Chaikin, P.~Grant, G.~J.~Morales, and P. Pincus,
\JCP {\bf 80}, 5776 (1984).

\bibitem{note4}
In ref.~\cite{Grier2}, a macroion valence of $Z=7300$ is obtained by
fitting the pair interaction between two isolated macroions with
a DLVO potential.  For bulk crystals, however, this valence somewhat
exceeds the approximate charge-renormalization limit~\cite{Alexander}.

\bibitem{Neu}
J.~C.~Neu, \PRL {\bf 82}, 1072 (1999).
 
\bibitem{Sader-Chan}
J.~E.~Sader and D.~Y.~C.~~Chan, \JCIS {\bf 213}, 268 (1999);
{\it Langmuir} {\bf 16}, 324 (2000).
 
\bibitem{Trizac}
E.~Trizac and J.~L.~Raimbault, \PR E {\bf 60}, 6530 (1999);
E.~Trizac, \PR E {\bf 62}, R1465 (2000).

\bibitem{Schmitz99}
K.~S.~Schmitz, {\it Phys.~Chem.~Chem.~Phys.} {\bf 1}, 2109 (1999).

\bibitem{Sear00}
R.~P.~Sear, \PR E {\bf 62}, 2501 (2000).

\bibitem{Russ02}
C.~Russ, H.~H.~von Gr\"unberg, M.~Dijkstra, R.~van Roij, \PR E {\bf 66},
011402 (2002).

\bibitem{Wu00}
J.~Z.~Wu, D.~Bratko, H.~W.~Blanch, and J.~M.~Prausnitz, \JCP {\bf 113},
3360 (2000).

\bibitem{Denton3}
A.~R.~Denton (unpublished).

\bibitem{Pethick70}
C.~J.~Pethick, \PR B {\bf 2}, 1789 (1970).

\bibitem{Brovman70}
E.~G.~Brovman and G.~Solt, {\it Solid State Comm.} {\bf 8}, 903 (1970).

\bibitem{Singh73}
S.~P.~Singh and W.~H.~Young, {\it J.~Phys.}~F: {\it Metal Phys.} {\bf 3}, 
1127 (1973).

\bibitem{Rasolt75}
M.~Rasolt and R.~Taylor, \PR B {\bf 11}, 2717 (1975);
L.~Dagens, M.~Rasolt, and R.~Taylor, \PR B {\bf 11}, 2726 (1975).

\bibitem{Louis98}
A.~A.~Louis and N.~W.~Ashcroft, \PRL {\bf 81}, 4456 (1998).

\bibitem{note5}
From Eq.~(\ref{appa-rho+r1}) on, to simplify notation, we dispense with 
angular brackets in denoting ensemble-averaged densities.



\end{references}
\end{document}